\documentclass[12pt,a4paper]{article}
\usepackage{graphicx}
\usepackage{amsmath}
\usepackage{cite}

\usepackage{color}

\def\Comment#1{}
\newcommand{\bean}{\begin{eqnarray*}}
\newcommand{\eean}{\end{eqnarray*}}

\newcommand{\gapproxeq}{\lower
.7ex\hbox{$\;\stackrel{\textstyle >}{\sim}\;$}}
\newcommand{\lapproxeq}{\lower
.7ex\hbox{$\;\stackrel{\textstyle <}{\sim}\;$}}

\newcommand\lsim{\mathrel{\rlap{\lower4pt\hbox{\hskip1pt$\sim$}}
    \raise1pt\hbox{$<$}}}
\newcommand\gsim{\mathrel{\rlap{\lower4pt\hbox{\hskip1pt$\sim$}}
    \raise1pt\hbox{$>$}}}
\newcommand{\ba}{\begin{array}}
\newcommand{\ea}{\end{array}}
\newcommand{\nn}{\nonumber}

\newcommand{\be}{\begin{equation}}
\newcommand{\ee}{\end{equation}}
\newcommand{\bear}{\begin{eqnarray}}
\newcommand{\eear}{\end{eqnarray}}

\newcommand{\ket}{\,\rangle}
\newcommand{\bra}{\langle \,}
\newcommand{\eqn}[1]{(\ref{#1})}
\newcommand{\cO}{{\cal O}}
\newcommand{\bel}[1]{\be\label{#1}}
\newcommand{\mL}{\mathcal{L}}

\newcommand{\mO}{\mathcal{O}}

\def\bat{\begin{array}{cc}}
\newcommand{\toG}{\stackrel{G}{\,\longrightarrow\,}}
\newcommand{\Frac}[2]{\frac{\displaystyle #1}{\displaystyle #2}}

\def\ie{{\it i.e.},\ }
\begin{document}
\thispagestyle{empty}
\begin{titlepage}
\begin{center}
\hfill BARI$-$TH/2012$-$653 \\
\hfill FTUV/12$-$0615 \\
\hfill IFIC/12$-$15

\vspace*{0.8cm}
\begin{Large}
{\bf
One-Loop Calculation of the Oblique S Parameter \\[10pt]
in Higgsless Electroweak Models}
 \\[1.3cm]
\end{Large}

{ \sc A. Pich$^{1}$ }, {\sc I. Rosell$^{1,2}$} and { \sc J.J. Sanz-Cillero$^3$ }  \\[0.8cm]

{\it $^{1}$ Departament de F\'\i sica Te\`orica, IFIC, Universitat de Val\`encia --
CSIC\\
 Apt. Correus 22085, E-46071 Val\`encia, Spain }\\[0.3cm]

{\it $^{2}$Departamento de Ciencias F\'\i sicas, Matem\'aticas y de la Computaci\'on, \\
Escuela Superior de Ense\~nanzas T\'ecnicas ESET, \\
Universidad CEU Cardenal Herrera, c/ Sant Bartomeu 55, \\
E-46115 Alfara del Patriarca, Val\`encia, Spain }\\[0.3cm]

{\it $^3$ Istituto Nazionale di Fisica Nucleare INFN, Sezione di Bari,\\
Via Orabona 4, I-70126 Bary, Italy  }\\[0.6cm]

\vspace*{0.9cm}
\begin{abstract}

\noindent
We present a one-loop calculation of the oblique $S$ parameter within Higgsless models
of electroweak symmetry breaking and analyze the phenomenological implications of the available electroweak precision data. We use the most general  effective Lagrangian with at most two
derivatives, implementing the chiral symmetry breaking $SU(2)_L\otimes SU(2)_R\to SU(2)_{L+R}$ with Goldstones, gauge bosons
and one multiplet of vector and axial-vector massive resonance states. Using the dispersive representation of Peskin and Takeuchi and imposing the short-distance constraints dictated by the operator product expansion, we obtain $S$ at the NLO in terms of a few resonance parameters.
In asymptotically-free gauge theories, the final result only depends on the vector-resonance mass
and requires $M_V > 1.8$~TeV (3.8 TeV) to satisfy the experimental limits at the 3$\sigma$ (1$\sigma$) level; the axial state is always
heavier, we obtain $M_A > 2.5$~TeV (6.6~TeV) at 3$\sigma$ (1$\sigma$). 
In strongly-coupled models, such as walking or conformal technicolour, where the second Weinberg sum rule does not apply, the vector and axial couplings are not determined by the short-distance constraints; but one can still derive a lower bound on $S$, provided the hierarchy $M_V < M_A$ remains valid. Even in this less constrained situation, we find that in order to satisfy the experimental limits at 3$\sigma$ one needs $M_{V,A} > 1.8$~TeV.
\end{abstract}
\end{center}
\vfill
\eject
\end{titlepage}

\pagenumbering{arabic}

\parskip12pt plus 1pt minus 1pt
\topsep0pt plus 1pt
\setcounter{totalnumber}{12}

\section{Introduction}

The Standard Model (SM) provides an extremely successful description of the
electroweak and strong interactions, which has been tested with high accuracy
in many experiments~\cite{EW-rev-Pich}.
A key feature of this theoretical framework is the particular mechanism adopted
to break the electroweak gauge symmetry $SU(2)_L\times U(1)_Y$
to the electromagnetic subgroup $U(1)_{\mathrm{QED}}$, so that the $W$ and $Z$ bosons become massive~\cite{HK-mechanism}. In order to generate the longitudinal polarizations of these spin-1 bosons (absent for massless gauge particles), one needs three additional degrees of freedom.
The SM implements the Electroweak Symmetry Breaking (EWSB), through an $SU(2)_L$ doublet of complex scalars $\Phi(x)$ and
a potential $V(\Phi)$ with non-trivial minima. The vacuum expectation value of the scalar
doublet generates the needed spontaneous symmetry breaking, giving rise to three Goldstone bosons which, in the unitary gauge, become the longitudinal polarizations of the gauge bosons.
Since $\Phi(x)$ contains four real fields, one massive neutral scalar survives in the physical spectrum: the Higgs boson. This particle is the main missing block of the SM.

The LHC has already excluded a broad range of Higgs masses, narrowing down the SM Higgs hunting to the low-mass region between 115.5 and 127 GeV
(95\% CL)~\cite{ATLAS-H:12,CMS-H:12}.\footnote{
The most recent, but still preliminary, data further restricts the allowed SM
Higgs masses to the range $\left[ 117.5 , 118.5 \right]\; \mathrm{GeV}\;\cup\; \left[ 122.5 , 127.5 \right]\; \mathrm{GeV}$ \cite{Moriond}.}
This is precisely the range of masses preferred by the global fit to precision electroweak data, which sets the upper bound $M_H< 169$~GeV at the 95\%
confidence level~\cite{Gfitter,LEPEWWG,ZFITTER}.
In the next months the LHC should find out whether such scalar field indeed exists. The discovery of a neutral boson in this mass range could provide a spectacular confirmation of the SM framework.

If the Higgs boson does not show up soon, we should look for alternative mechanisms of mass generation, satisfying the many experimental constraints which the SM has successfully fulfilled so far. Actually, the existing phenomenological tests
have only confirmed the pattern of symmetry breaking, but not the detailed dynamics embodied in the Higgs potential.
The scalar sector of the SM Lagrangian can be written in the form
\cite{Pich:1995bw,Pich:1998xt}
\bel{eq:l_sm}
\mL(\Phi)\, =\, {1\over 2}\, \langle\, \left(D^\mu\Sigma\right)^\dagger D_\mu\Sigma\,\rangle
- {\lambda\over 16} \left(\langle\,\Sigma^\dagger\Sigma\,\rangle
- v^2\right)^2 ,
\ee
where
\be
\Sigma \,\equiv\, \left( \Phi^c, \Phi\right) \, =\,
\left( \bat \Phi^{0*} & \Phi^+  \\ -\Phi^- &  \Phi^0 \ea\right)\, ,
\label{eq:sigma_matrix}
\ee
$D_\mu\Sigma \equiv \partial_\mu\Sigma
+ i g \,\frac{\vec{\sigma}}{2}\vec{W}_\mu \,\Sigma - i g' \,\Sigma \,\frac{\sigma_3}{2} B_\mu $
is the usual gauge-covariant derivative and $\langle A\rangle$ stands for the trace of the $2\times 2$ matrix $A$.
In the limit where the $U(1)_Y$ coupling $g'$ is neglected,
$\mL(\Phi)$ is invariant
under global $G\equiv SU(2)_L\otimes SU(2)_R$ transformations
\be
\Sigma \, \toG \,
g_L \,\Sigma\, g_R^\dagger , \qquad\qquad\qquad
g_{L,R}  \in SU(2)_{L,R} \, .
\label{eq:sigma_transf}
\ee
Thus, the scalar sector has an additional global $SU(2)_R$ symmetry; only its
$U(1)_Y$ subgroup is gauged in the SM.
Performing a polar decomposition,
\be
\Sigma(x) \, = \, {1\over\sqrt{2}}
\left[ v + H(x) \right] \, U(\varphi(x)) \, , \qquad\qquad\qquad
U(\varphi) \, =\,  \exp{\left\{ i \vec{\sigma} \,
\vec{\varphi} / v \right\} } \, ,
\label{eq:polar}
\ee
in terms of the Higgs field $H(x)$ and the Goldstones
$\vec{\varphi}(x)$,
and taking the limit $\lambda\gg 1$ (heavy Higgs),
we can rewrite $\mL(\Phi)$ in the form \cite{AB:80}:
\be
\mL(\Phi)\, =\, {v^2\over 4}\,
\langle\, D_\mu U^\dagger D^\mu U \,\rangle \, +\,
\cO\left( H/ v \right) ,
\label{eq:sm_goldstones}
\ee
with
$D_\mu U \equiv \partial_\mu U
+ i g \,\frac{\vec{\sigma}}{2}\vec{W}_\mu \, U - i g'\, U \,\frac{\sigma_3}{2} B_\mu$.
In the unitary gauge $U=1$, this Lagrangian
reduces to the usual bilinear gauge-mass term, with
$Z^\mu \equiv \cos{\theta_W} W_3^\mu - \sin{\theta_W} B^\mu$,
$M_W = M_Z \,\cos{\theta_W} = v g/2$ and
$\tan{\theta_W} = g'/ g$.

The term proportional to $v^2$ in Eq.~\eqn{eq:sm_goldstones} is the
universal model-independent lowest-order Goldstone Lagrangian associated with the
symmetry breaking $SU(2)_L\otimes SU(2)_R\to SU(2)_{L+R}$. In Quantum Chromodynamics (QCD)
this Lagrangian describes the dynamics of pions at $\cO (p^2)$
(two derivatives), with $v = f_\pi$, the pion decay constant \cite{Pich:1995bw}.
The same Lagrangian
with $v = \left(\sqrt{2} G_F\right)^{-1/2} = 246\:\mathrm{GeV}$
describes the Goldstone boson dynamics associated with the EWSB.
The electroweak global $SU(2)_{L+R}$ is usually called custodial symmetry group
\cite{Sikivie:1980hm}.

In the absence of direct compelling evidence for a light Higgs boson, one should investigate the implications of the assumed Goldstone symmetry structure, independently of any particular implementation of the symmetry breaking. This can be done applying the same momentum expansion techniques used in Chiral Perturbation Theory ($\chi$PT)
to describe low-energy QCD \cite{Pich:1995bw,ChPT,ChPTp4,Ecker:1994gg}.
The electroweak Goldstone dynamics is then parameterized through an Effective Lagrangian
which contains the SM gauge symmetry realized nonlinearly
\cite{Appelquist:1980ix,Appelquist:1980vg,Longhitano:1980iz,Dobado:1990zh,Dobado:1997jx}.
Only the known light degrees
of freedom (leptons, quarks and gauge bosons) appear in this effective Lagrangian, which does not include any Higgs field.
With an appropriate choice of its parameters, the electroweak chiral Lagrangian includes the SM as long as the energies involved are small compared with the Higgs mass. In addition, it can also accommodate any model that reduces to the SM at low energies; in particular,
Higgsless electroweak models.

In strongly-coupled models
the gauge symmetry is dynamically broken by means of some non-perturbative interaction.
Usually, theories of this kind do not contain any fundamental Higgs, bringing instead resonances of different types as happens in QCD
\cite{Chivukula:1998if,Pomarol:2012sb,Andersen:2011yj}.
For instance, Technicolour \cite{technicolor},
the most studied strongly-coupled model, introduces an asymptotically-free QCD replica at TeV energies which breaks the electroweak symmetry in the infrared,
in a similar way as chiral symmetry is broken in QCD.
This gives rise to the appearance of a tower of heavy resonances
in the scattering amplitudes.
Other models consider the possibility that the ultraviolet (UV) theory remains close to a
strongly-interacting conformal fixed point over a wide range of energies
(Walking Technicolour) \cite{walking}; recent work in this direction
incorporates conformal field theory techniques (Conformal Technicolour)
\cite{Luty:2004ye,Rattazzi:2008pe,new}.
Most of the recent activity has focused on strongly-coupled models in
warped \cite{Randall:1999ee} or deconstructed \cite{ArkaniHamed:2001ca} extra dimensions~\cite{Csaki:2003zu,SekharChivukula:2001hz,Agashe:2003zs,Foadi:2003xa}.

In this paper we reanalyze the main constraint from electroweak precision tests
on strongly-coupled models: the  oblique $S$ parameter \cite{Peskin:92}. We perform a
one-loop calculation of this important quantity within an effective low-energy theory including the electroweak Goldstones and resonance fields.
The theoretical framework is completely analogous to the Resonance Chiral Theory
(R$\chi$T) description of QCD at GeV energies \cite{RChT,Pich:2002xy}.
In recent years a thorough investigation of R$\chi$T at the one-loop level has been performed \cite{Cata:2001nz,L9a,L8,L10,L9,Natxo-thesis,RPP:05,RChT-EoM},
bringing an improved understanding of the resonance dynamics.
We can profit the available QCD results to investigate similar issues in the electroweak sector. In particular, we will make use of the procedure developed to compute
the low-energy constants of $\chi$PT
at the next-to-leading order (NLO) through a matching with R$\chi$T~\cite{Cata:2001nz,L9a,L8,L10,L9}.
The estimation of $S$ in strongly-coupled electroweak models is equivalent
to the calculation of $L_{10}$ in $\chi$PT~\cite{L10}.
We only need to translate the results of Ref.~\cite{L10} to the electroweak context.

Several one-loop estimates of the electroweak $S$ and $T$ parameters in
the three-site \cite{Matsuzaki:2006wn} and more general
\cite{S-Isidori:08,S-Cata:10,S-Orgogozo:11}
Higgless models have appeared recently. The results of
Refs.~\cite{Matsuzaki:2006wn,S-Isidori:08,S-Cata:10} contain an unphysical dependence on the UV cut-off,
which manifests the need for local contributions to account for a proper UV
completion. A slightly more general Lagrangian has been considered in Ref.~\cite{S-Orgogozo:11} which, moreover, takes advantage of the dispersive approach
suggested in \cite{Peskin:92} to soften the UV problem.
As shown in Refs.~\cite{L8,L10,L9}, the dispersive approach
avoids all technicalities associated with the renormalization procedure,
allowing us to understand the underlying physics in a much more transparent way.
A crucial ingredient of this approach is the assumed UV behaviour of the
relevant Green functions.

We will closely follow the dispersive approach of Ref.~\cite{L10}
with a general resonance Lagrangian with at most two derivatives.
In Section~\ref{sec.observables}, we briefly review the definition of
the $S$ and $T$ parameters and the dispersive representation of $S$ advocated
by Peskin and Takeuchi \cite{Peskin:92}.
The effective electroweak Lagrangian, including the lightest vector and axial-vector resonances,
is constructed in Section~\ref{sec.lagrangian}.
We compute next the resonance contributions to the relevant spectral functions, at the one-loop level; the NLO contributions to the $S$ parameter are discussed in
Section~\ref{sec.NLO} and we study a series of high-energy constraints in Section~\ref{sec.SD-constraints}.
The phenomenological outcomes from the various available constraints
are compared to the experimental data in Section~\ref{sec.pheno},
where we establish the present lower bounds on the vector and axial-vector resonance masses.
Our conclusions are finally summarized in Section~\ref{sec.conclusions}.
Some technical aspects related with the dispersive integration and
the precise expression of the one-loop spectral function are given in the Appendixes.

\section{Oblique electroweak observables}
\label{sec.observables}

We focus our study on the universal {\it oblique}
corrections that occur via the electroweak boson self-energies.
The computation is performed in the Landau gauge, so that the gauge boson propagators
are transverse and their self-energies,
\begin{equation}
\mathcal{L}_{\mathrm{v.p.}}\,\dot=\,
- \frac{1}{2} W^3_\mu\, \Pi^{\mu\nu}_{33}(q^2)W^3_\nu
-\frac{1}{2}B_\mu\,\Pi^{\mu\nu}_{00}(q^2) B_\nu
- W^3_\mu\, \Pi^{\mu\nu}_{30}(q^2) B_\nu
- W^+_\mu\, \Pi^{\mu\nu}_{WW}(q^2)W^-_\nu\,,
\end{equation}
can be decomposed as
\begin{equation}
\Pi^{\mu\nu}_{ij} (q^2) \,=\, \left( -g^{\mu\nu} +
\frac{q^\mu q^\nu}{q^2}\right)\; \Pi_{ij}(q^2)\, .
\end{equation}

The vacuum polarization amplitudes are expected to contain the dominant contributions
from new physics beyond the SM. When the masses of the new particles are much larger than $M_Z$, it is useful to perform a series expansion in powers of $M_Z/M_{\mathrm{new}}$
\cite{Peskin:92,Kennedy:1990ib,S-def_Barbieri,S-def_Pomarol}. Most of the effects on precision electroweak measurements can be described in terms of three parameters $S$, $T$ and $U$ (or equivalently $\varepsilon_1$, $\varepsilon_2$ and $\varepsilon_3$).
$S$ ($S+U$) parameterizes the new-physics contributions to the difference between the
$Z$ ($W$) self-energy at $Q^2 = M_Z^2$ ($Q^2=M_W^2$) and $Q^2=0$, while $T$ is
proportional to the difference between the new-physics contributions to the $W$ and $Z$
self-energies at $Q^2=0$. Most simple types of new physics give $U=0$, which we will not discuss any further. The precise definitions of $S$ and $T$ involve the quantities
\begin{equation}
e_3\,=\, \Frac{g}{g'}  \; \widetilde{\Pi}_{30}(0) \, ,
\qquad\qquad
e_1\,=\,
\frac{ \Pi_{33}(0) - \Pi_{WW} (0)}{M_W^2}\,,
\label{eq.S-def}
\end{equation}
where the tree-level Goldstone contribution has been removed from $\Pi_{30}(q^2)$
in the form \cite{Peskin:92}:
\begin{equation}
\Pi_{30}(q^2)\,=\,q^2\, \widetilde\Pi_{30}(q^2)\,+\,\frac{g^2 \tan{\theta_W}}{4}\, v^2 \,  .
\label{eq.PiTilde}
\end{equation}
The $S$ and $T$ parameters are given by the deviation
with respect to the SM contributions  $e_3^{\rm SM}$ and $e_1^{\rm SM}$,
respectively:
\begin{equation}
S\,=\,  \Frac{16\pi}{g^2}\;\big(e_3 - e_3^{\rm SM}\big)\, ,
\qquad \qquad
T\,=\, \Frac{4\pi \sin^2{\theta_W}}{g^2}\; \big(e_1-e_1^{\rm SM}\big)  \,.
\end{equation}

In this paper we will concentrate on the $S$ parameter, for which a useful
dispersive representation was  introduced by
Peskin and Takeuchi~\cite{Peskin:92}:
\begin{eqnarray}
S  &=&   \frac{16}{g^2  \tan{\theta_W}}\; \int_0^\infty
\frac{\mathrm{d} s}{s}\;
\bigg(
\mathrm{Im}  \widetilde{\Pi}_{30}(s) - \mbox{Im}\widetilde{\Pi}_{30}^{\rm SM}(s)
\bigg)
\,=\, \nonumber \\[5pt]
&=&  \int_0^\infty
 \frac{\mathrm{d} s}{s} \;\,\left( \frac{16}{g^2  \tan{\theta_W}}\;
  \mathrm{Im}  \widetilde{\Pi}_{30}(s)
  -    \frac{1}{12\pi}\left[ 1
  - \left(1-\frac{M_H^2}{s} \right)^3 \theta \left(s-M_H^2 \right) \right] \right) \,  .
\label{eq.S-dispersive}
\end{eqnarray}
Note that in order to define the SM contribution, and therefore $S$, one needs a reference value for the SM Higgs mass.
The convergence of this unsubtracted dispersion relation requires a vanishing
spectral function at short distances.
In the SM, $\mathrm{Im}\widetilde{\Pi}_{30}(s)$ vanishes at $s\to\infty$
due to the interplay of the
two-Goldstone and  the Goldstone--Higgs contributions.
In the absence of a Higgs boson, this UV convergence will be realized in a different way.

Sum rules of this type have been widely used in QCD
for the extraction of $\chi$PT low-energy constants from
experimental data~\cite{L10-sum-rule}.   They have also provided
successful  determinations when applied to theoretical computations
in R$\chi$T~\cite{L8,L10,L9}.

\section{Electroweak effective theory}
\label{sec.lagrangian}

Let us consider a low-energy effective theory containing
the SM gauge bosons coupled to the electroweak Goldstones
(we will not discuss the fermion couplings\footnote{
A recent discussion of fermion operators in the effective Goldstone electroweak theory,
with references to previous work, can be found in Ref.~\cite{Buchalla:2012qq}.
}).
We will only assume the SM pattern of EWSB, \ie that the
theory is symmetric under $G=SU(2)_L\otimes SU(2)_R$
and becomes spontaneously broken to the diagonal subgroup $H=SU(2)_{L+R}$.
The Lagrangian can be organized as an expansion in powers of derivatives
(momenta) over the EWSB scale. At the lowest order, the Lagrangian takes
the form
\begin{equation}
\mathcal{L}_{\mathrm{EW}}^{(2)} \; =\;
-\frac{1}{2g^2}\,\bra \hat{W}_{\mu\nu} \hat{W}^{\mu\nu} \ket
\, -\,\frac{1}{2g'^{\, 2}}\,\bra  \hat{B}_{\mu\nu} \hat{B}^{\mu\nu} \ket
\, +\, \frac{v^2}{4}\, \bra u_\mu u^\mu \ket \, ,
\end{equation}
which contains the usual Yang-Mills terms plus the Goldstone interactions
in Eq.~(\ref{eq:sm_goldstones}). We have used the notation
\begin{eqnarray}
\hat{W}^{\mu\nu}\, =\, \partial^\mu \hat{W}^\nu  - \partial^\nu \hat{W}^\mu - i \, [\hat{W}^\mu,\hat{W}^\nu]\, ,
\qquad\qquad
\hat{B}^{\mu\nu}\, =\, \partial^\mu \hat{B}^\nu - \partial^\nu \hat{B}^\mu
- i \, [\hat{B}^\mu,\hat{B}^\nu]\, ,
\nn\\
u^\mu = i\, u\,  D^\mu U^\dagger\, u = -i\, u^\dagger  D^\mu U\, u^\dagger = u^{\mu\dagger}\, ,
\qquad\qquad\;
D^\mu U = \partial^\mu U - i \, \hat{W}^\mu U+ i \,U\, \hat{B}^\mu\, .\;\;
\end{eqnarray}
The Goldstone bosons are parameterized through
$U=u^2=\exp{\left\{ i \vec{\sigma} \vec{\varphi} / v \right\} }$,
where $u(\varphi)$ is an element of the coset $G/H$.
Under a transformation\ $g\equiv (g_L,g_R)\in G$,\footnote{
For a given choice of coset representative $\bar\xi(\varphi)\equiv \left(\xi_L(\varphi),\xi_R(\varphi)\right)\in G$, the change of the
Goldstone coordinates under a chiral transformation takes the form
$$ \xi_L(\varphi) \to g_L\, \xi_L(\varphi)\, h^\dagger(\varphi,g)\, ,
\qquad\qquad\qquad
\xi_R(\varphi) \to g_R\, \xi_R(\varphi)\, h^\dagger(\varphi,g)\, . $$
The same compensating transformation $h(\varphi,g)$ occurs in both chiral sectors because
they are related by a discrete parity transformation $L\leftrightarrow R$ which leaves
$H$ ($L+R$) invariant. $U(\varphi)\equiv \xi_L(\varphi)\xi_R^\dagger(\varphi)$ transforms
as $g_L\, U(\varphi)\,g_R^\dagger$.
We take a canonical choice of coset
representative such that $\xi_L(\varphi) = \xi_R^\dagger(\varphi)\equiv u(\varphi)$.
}
%
\be
u(\varphi)\quad \longrightarrow\quad g_L \, u(\varphi)\, h^\dagger(\varphi,g)\,=\,  h(\varphi,g) \, u(\varphi)\,  g_R^\dagger\, ,
\ee
with $h\equiv h(\varphi,g)\in H$ a compensating transformation to
preserve the coset representative \cite{Coleman:1969sm}.
Requiring the $SU(2)$ matrices $\hat{W}^\mu$ and $\hat{B}^\mu$ to transform as
\bel{eq:FakeTransform}
\hat{W}^\mu\to g_L\, \hat{W}^\mu g_L^\dagger + i\, g_L\, \partial^\mu g_L^\dagger\, ,
\qquad\qquad
\hat{B}^\mu\to g_R\, \hat{B}^\mu g_R^\dagger + i\, g_R\, \partial^\mu g_R^\dagger\, ,
\ee
the effective Lagrangian is invariant under local $SU(2)_L\otimes SU(2)_R$
transformations. The identification
\bel{eq:SMgauge}
\hat{W}^\mu \, =\, -g\;\frac{\vec{\sigma}}{2}\, \vec{W}^\mu \, ,
\qquad\qquad\qquad
\hat{B}^\mu\, =\, -g'\;\frac{\sigma_3}{2}\, B^\mu\, ,
\ee
breaks explicitly the $SU(2)_R$ symmetry group, in exactly the same way as the
SM does, preserving the $SU(2)_L\otimes U(1)_Y$ gauge symmetry.
Taking functional derivatives with respect to the formal left and right sources $\hat{W}^\mu$ and $\hat{B}^\mu$, one can also study the corresponding currents (and current Green functions).

The inner nature of the EWSB is left unspecified. Instead of the SM Higgs,
we assume that the strongly-coupled underlying dynamics gives rise to massive resonance multiplets transforming as triplets ($R\equiv \frac{\vec{\sigma}}{2}\, \vec{R}$) or singlets ($R_1$) under $H$:
\be
R\quad \longrightarrow\quad h(\varphi,g) \: R\: h^\dagger(\varphi,g)\, ,
\qquad\qquad\qquad
R_1\quad \longrightarrow\quad R_1\, .
\ee
In order to build invariant operators under the assumed symmetry group,
it is useful to introduce \cite{RChT} the covariant derivative
\be
\nabla^\mu R \, =\, \partial^\mu R +  \left[ \Gamma^\mu , R\right]\, ,
\qquad\qquad
\Gamma^\mu \,=\, \Frac{1}{2}  \left\{u \left( \partial^\mu
- i \hat{B}^\mu \right) u^\dagger +u^\dagger \left(\partial^\mu - i \hat{W}^\mu \right) u \right\}
\, ,
\ee
and
\be
h^{\mu\nu}\, =\, \nabla^\mu u^\nu + \nabla^\nu u^\mu \, ,
\qquad\qquad
f^{\mu\nu}_{\pm} \, =\,
u^\dagger \,\hat{W}^{\mu\nu} u \pm  u \,\hat{B}^{\mu\nu} u^\dagger \, ,
\ee
which transform as triplets under $H$:
\be
\left\{ \nabla^\mu R\, ,\, h^{\mu\nu}\, ,\, f^{\mu\nu}_{\pm}\, ,\, u^\mu\right\}
\quad \longrightarrow\quad h \; \left\{ \nabla^\mu R\, ,\, h^{\mu\nu}\, ,\, f^{\mu\nu}_{\pm}\, ,\, u^\mu\right\}\; h^\dagger\, .
\ee

We must add to the effective Lagrangian the couplings of the resonances
to the electroweak Goldstones, preserving the invariance under $G$ transformations.
To simplify the discussion we will only consider the lightest multiplets, which are expected
to be the most relevant ones at low energies.
An heuristic rule in QCD along the past years have been to construct the resonance
Lagrangian with operators with the lowest number of
derivatives~\cite{L10,RChT,Natxo-thesis,RPP:05},  typically $\cO(p^2)$,
as terms with higher powers of momenta tend to violate the
expected behaviour of the QCD matrix elements at high energies.
In other cases, one may prove that a given kind of operators
can be always simplified to the simplest resonance operator
and others with only Goldstones by means of convenient field redefinitions
and the equations of motion~\cite{RChT-EoM}.
Given our ignorance about the underlying EWSB dynamics, we will adopt the
same procedure and will restrict the Lagrangian to low powers of momenta in order to guarantee a good UV behaviour.
The Lagrangian can be then classified  accordingly to the number
of resonance fields. In the QCD case, a complete set of operators with one, two and three resonance fields has been given in Refs.~\cite{RChT}, including vector, axial-vector, scalar
and pseudoscalar resonances. These results can be easily adapted to the electroweak
effective theory with very simple notational changes.

For instance, adding the most general couplings of a singlet scalar one gets an effective
theory which includes as a particular case the SM Higgs potential \cite{Bagger:1993zf}.
This effective theory allows to describe more general possibilities such as the
composite-Higgs scenario \cite{Contino:2010mh}
where the light scalar emerges as a pseudo-Goldstone boson from a
strongly-coupled sector \cite{Kaplan:1983fs}.
The phenomenological implications of this effective Lagrangian are being actively
investigated at present \cite{Grober:2010yv}.

In this work we are interested in the lowest-mass vector ($V^{\mu\nu}$)
and axial-vector ($A^{\mu\nu}$) resonances,
which can induce sizeable corrections to the gauge-boson self-energies.
We will use the antisymmetric tensor formalism\footnote{
In addition to provide the same type of description for vector and axial-vector states,
this formalism avoids the mixing of the axial resonances with the Goldstones and
its softer momentum dependence allows us to recover in a simpler way the
right UV behaviour.}
to describe these spin--1 fields \cite{ChPTp4,RChT}
and will assume that the strong dynamics
preserves parity ($L\leftrightarrow R$) and charge conjugation. The corresponding
Lagrangians can then be directly taken from Refs.~\cite{RChT}. We will need operators
with one resonance field,
\be
\mathcal{L}_V\, +\, \mathcal{L}_A\; =\;
\frac{F_V}{2\sqrt{2}}\, \bra V_{\mu\nu} f^{\mu\nu}_+ \ket \,
+\, \frac{i\, G_V}{2\sqrt{2}}\, \bra V_{\mu\nu} [u^\mu, u^\nu] \ket
\, + \, \frac{F_A}{2\sqrt{2}}\, \bra A_{\mu\nu} f^{\mu\nu}_- \ket\, ,
\ee
and the following terms with two resonances:
\begin{eqnarray}
\mathcal{L}^{\,\mathrm{kin}}_{RR} &=&
-\frac{1}{2}\, \bra \nabla^\lambda R_{\lambda\mu} \nabla_\nu R^{\nu\mu}\,-\,
\frac{M_R^2}{2}\, R_{\mu\nu}R^{\mu\nu} \ket \, ,
\qquad\qquad (R=V,A)    \\
\mathcal{L}_{VA}&=&\,
i \,\lambda^{VA}_2\, \bra  [ V^{\mu\nu}, A_{\nu\alpha} ] h^\alpha_\mu \ket \,+\, i\, \lambda^{VA}_3 \,\bra  [ \nabla^\mu V_{\mu\nu}, A^{\nu\alpha} ] u_\alpha \ket\phantom{\frac{1}{2}}
\nonumber \\ &&
+\, i \,\lambda^{VA}_4 \,\bra  [ \nabla_\alpha V_{\mu\nu}, A^{\alpha\nu} ] u^\mu \ket
\, +\, i \,\lambda^{VA}_5\, \bra  [ \nabla_\alpha V_{\mu\nu}, A^{\mu\nu} ] u^\alpha \ket \phantom{\frac{1}{2}}
\nonumber \\ &&
+\, i \,\lambda^{VA}_6\, \bra  [
V_{\mu\nu}, A^{\mu}_{\,\,\alpha} ]f^{\alpha\nu}_- \ket  \phantom{\frac{1}{2}} \, .
\label{VALagrangian}  
\end{eqnarray}
The calculation of the $S$ parameter does not actually depend on the
five $\mL_{VA}$ couplings but only on two particular combinations of them:
\begin{equation}\label{eq.kappa}
\kappa\,=\, -2 \lambda_2^{\mathrm{VA}} + \lambda_3^{\mathrm{VA}} \, ,
\qquad\qquad\qquad
\sigma \,=\,   2 \lambda_2^{\mathrm{VA}} -2 \lambda_3^{\mathrm{VA}}
+ \lambda_4^{\mathrm{VA}} + 2 \lambda_5^{\mathrm{VA}} \,.
\end{equation}
In principle, one might build operators with two vector or axial-vector resonances ($\mathcal{L}_{VV}$ and $\mathcal{L}_{AA}$) or more than two resonance fields~\cite{L10,RChT,Natxo-thesis,RPP:05} but, since we will neglect the contributions from two-resonance absorptive cuts, they will not play any role in the present work.

Collecting all pieces, the effective Lagrangian we are going to use reads
\be
\mathcal{L} \, =\, \mathcal{L}_{\mathrm{EW}}^{(2)} + \mathcal{L}_{\mathrm{GF}}
+\mathcal{L}_{V}+\mathcal{L}_{A} +\mathcal{L}_{VV}^{\mathrm{kin}} + \mathcal{L}_{AA}^{\mathrm{kin}}   +\mathcal{L}_{VA}\, ,
\ee
with
\be
\mathcal{L}_{\mathrm{GF}}\, =\, -\frac{1}{2\xi}\, (\partial^\mu \vec{W}_\mu)^2
\ee
the gauge-fixing term.
The calculation of the oblique parameter $S$ will be performed in the Landau gauge
$\xi = 0$.
This eliminates any possible mixing of the Goldstones and the gauge bosons,
which can only occur through the longitudinal parts of the $W^\pm$ and $Z$ propagators.

\section{NLO calculation of $S$}
\label{sec.NLO}

\begin{figure}
\begin{center}
\includegraphics[scale=0.8]{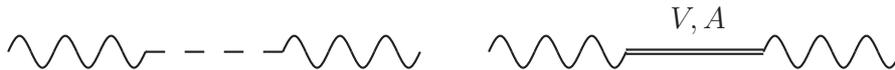}
\caption{\small{Leading-order contributions to
$\Pi_{30}(s)$. A dashed line stands for a Goldstone boson, a double line indicates a resonance field and a curved line represents a gauge boson.}} \label{LO_graphs}
\end{center}
\end{figure}

The tree-level contributions to the gauge-boson vacuum polarization $\Pi_{30}(s)$
are shown in Figure~\ref{LO_graphs} and lead to the well-known
leading-order (LO) result
\cite{Peskin:92,S-Isidori:08,S-Cata:10,S-Orgogozo:11}:
\begin{equation}
\left. \Pi_{30}(s) \right|_{\mathrm{LO}}\; =\;
\frac{g^2  \tan{\theta_W} }{4}\; s\;  \left(\frac{v^2}{s}+  \frac{F_V^2}{M_V^2-s}
- \frac{F_A^2}{M_A^2-s} \right)\, .
\label{eq.TL}
\end{equation}
The first term contains the Goldstone pole, which determines $\Pi_{30}(0)$.
This constant piece (also present in the SM) has been subtracted in
the definition of $\widetilde\Pi_{30}(s)$ in
Eqs.~(\ref{eq.S-def}) and (\ref{eq.PiTilde})
and does not play any role in the $S$ parameter:
\be\label{eq.S_LO}
S_{\mathrm{LO}} \,=\, 4\pi\, \left( \frac{F_V^2}{M_V^2} - \frac{F_A^2}{M_A^2} \right) \, .
\ee
The result can be trivially generalized to incorporate the exchange
of several vector and axial-vector resonance multiplets.

Notice that the experimental value of $S$ is provided at a given reference value
of the Higgs mass. However, the SM Higgs contribution only appears at the one-loop level.
Thus, there is a scale ambiguity when comparing the leading-order
theoretical result with the experimental constraint. This is similar to what
happens in QCD with the tree-level estimate of the analogous parameter $L_{10}$,
which does not capture its renormalization-scale dependence.
In both cases, a one-loop calculation is needed to fix the ambiguity.

The NLO contribution is most efficiently obtained through a dispersive calculation.
The essential condition needed to properly define the Peskin-Takeuchi representation
in Eq.~(\ref{eq.S-dispersive}) is a vanishing spectral function
$\mathrm{Im}\widetilde{\Pi}_{30}(s)$ at $s\to\infty$; \ie
the correlator $\Pi_{30}(s)$ should behave at most as a constant at short distances.
This allows us to reconstruct the correlator from the spectral function:
\be\label{eq.dispersivePi}
\Pi_{30}(s)\,= \, \Pi_{30}(0)\, +\, \Frac{s}{\pi}\,\int_0^\infty \, \Frac{\mathrm{d}t}{t\,(t-s)}\,
\mbox{Im}\Pi_{30}(t)\, .
\ee
Some care has to be taken with the simultaneous presence of resonance poles and two-particle
cuts. For simplicity, we omit here all technical aspects concerning the dispersive integral and the integration circuit. A more precise discussion is given in Appendix~\ref{app.dispersive}.

\begin{figure}
\begin{center}
\includegraphics[scale=0.5]{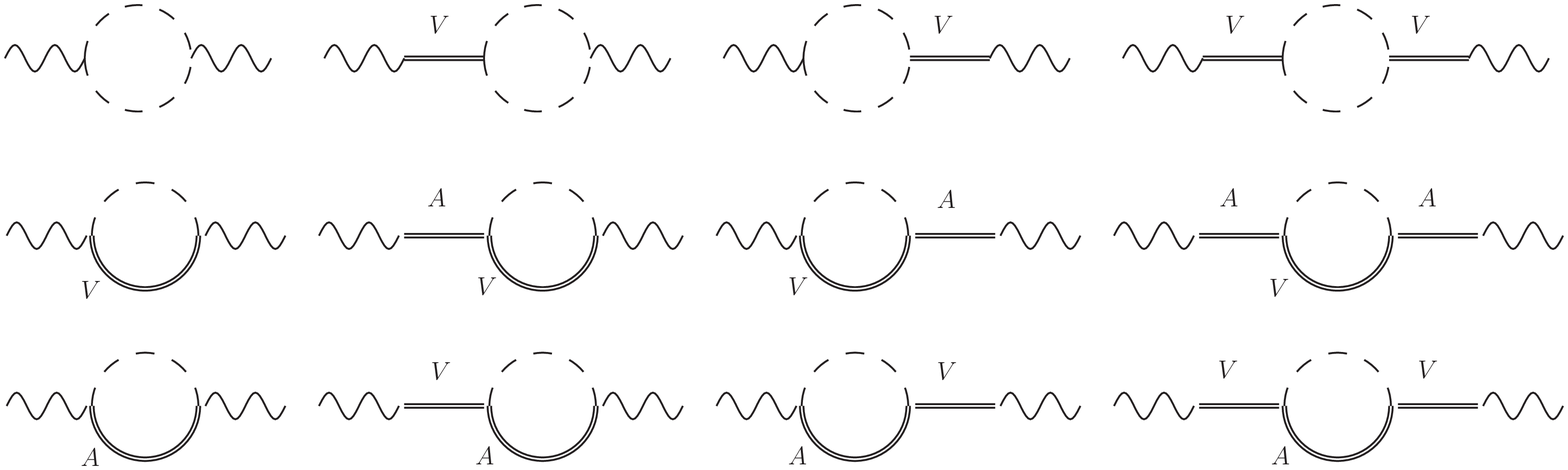}
\caption{\small{NLO contributions to $\mbox{Im} \Pi_{30}(s)$. A dashed line stands for a Goldstone boson, a double line indicates a resonance field and a curved line represents a gauge boson.}}
\label{NLO_graphs}
\end{center}
\end{figure}

Figure \ref{NLO_graphs} shows the one-loop contributions to $\Pi_{30}(s)$ generating
absorptive parts. We have considered two-particle cuts with two Goldstones or one Goldstone
plus one massive resonance, either vector or axial-vector. The two Goldstone contribution
is also present in the SM and, therefore, cancels out from the $S$ parameter; this guarantees the good infrared behaviour of the representation (\ref{eq.S-dispersive}).
We neglect the absorptive contributions from cuts with two resonances,
which are kinematically suppressed by their much heavier thresholds.
The explicit results for the different spectral functions are given in Appendix~\ref{app.spectralfunctions}.
Using the once-subtracted dispersion relation  for $\Pi_{30}(s)$,
the total NLO result, including the tree-level exchanges, can be written in
the form~\cite{L10,Natxo-thesis}
\be
\left. \Pi_{30}(s) \right|_{\mathrm{NLO}}  \, =\,
\frac{g^2\tan{\theta_W} }{4}\;  s \;  \left(\frac{v^2}{s}+  \frac{F_{V}^{r\,2}}{M_{V}^{r\,2}-s}
- \frac{F_{A}^{r\,2}}{M_{A}^{r\,2}-s} \; +\; \overline{\Pi}(s)\right)\, ,
\label{eq.T-NLO}
\ee
where $F_R^r$ and $M_R^r$ are ``renormalized'' couplings which properly
define the resonance poles at the one-loop level.
The one-loop contribution from the two-particle cuts is provided by $\overline{\Pi}(s)$. The $S$ parameter is given by
\be\label{eq.Sbar}
S_{\mathrm{NLO}} \,=\,
4\pi \left( \frac{F_{V}^{r\,2}}{M_{V}^{r\,2}} - \frac{F_{A}^{r\,2}}{M_{A}^{r\,2}} \right)
\; +\; \overline{S} \, .
\ee
The precise definitions of $\overline{\Pi}(s)$
and $\overline{S}$ are given in Appendix~\ref{app.dispersive}.

\section{High-energy constraints}
\label{sec.SD-constraints}

The two-particle spectral functions are determined by seven parameters:
$F_V$, $F_A$, $G_V$, $\kappa$, $\sigma$, $M_V$ and $M_A$.
The number of unknown couplings can be reduced using short-distance information
\cite{Pich:2002xy}. However, in contrast with the QCD case, we ignore here the underlying dynamical theory. We explore next the various high-energy constraints which can be
considered for the extraction of $S$. 

\subsection{Weinberg sum rules}

Since we are assuming that weak isospin and parity are good symmetries of the strong
dynamics, the correlator $\Pi_{30}(s)$ can be written in terms of the vector ($R+L$) and axial-vector ($R-L$) two-point functions as \cite{Peskin:92}
\be
\Pi_{30}(s)\, =\, \frac{g^2 \tan{\theta_W}}{4}\; s\;
\left[ \Pi_{VV}(s) - \Pi_{AA}(s)\right]\, .
\ee
The short-distance behaviour of this difference can be analyzed through the
Operator Product Expansion (OPE) of the right and left currents.
Owing to the chiral symmetry of the underlying theory, the only non-zero contributions
involve order parameters of the EWSB, \ie operators invariant under $H$ but not under $G$.
This guarantees the convergence of the dispersion relation (\ref{eq.dispersivePi})
because the unit operator is obviously symmetric.
In asymptotically-free gauge theories the difference $\Pi_{VV}(s)-\Pi_{AA}(s)$ vanishes at $s\to\infty$ as $1/s^3$ \cite{Bernard:1975cd}. This implies two super-convergent sum rules,
known as the first and second Weinberg sum rules (WSRs)~\cite{WSR}:
\begin{eqnarray}
\frac{1}{\pi}\,\int_0^\infty\;\mathrm{d}t\;
\left[\mathrm{Im}\Pi_{VV}(t)-\mathrm{Im}\Pi_{AA}(t)\right] &=& v^2\, ,
\\[10pt]
\frac{1}{\pi}\,\int_0^\infty\; \mathrm{d}t\; t\;
\left[\mathrm{Im}\Pi_{VV}(t)-\mathrm{Im}\Pi_{AA}(t)\right] &=& 0\, .
\end{eqnarray}
It is likely that the first of these sum rules is also true in gauge theories
with non-trivial UV fixed points.\footnote{
The specific condition required is that the OPE of $\Pi_{VV}(s)-\Pi_{AA}(s)$ 
does not contain operators with physical scaling dimension as low as 4
(for the second sum rule) or 2 (for the first)  \cite{Peskin:92}.}
However, the second WSR cannot be used in Conformal Technicolour models \cite{S-Orgogozo:11}
and its validity is questionable in most Walking Technicolour scenarios \cite{Appelquist:1998xf}.

\subsubsection{WSR constraints at leading order}

From the short-distance expansion of Eq.~(\ref{eq.TL}), one easily obtains the
implications of the WSRs at LO. The first WSR imposes the relation
\begin{equation}
 F_{V}^2 \,-\, F_{A}^2 \, =\, v^2 \, ,
\label{eq:1stWSR-LO}
\end{equation}
while requiring $ \Pi_{30}(s)$ to vanish as $1/s^2$  at short distances
(second WSR) leads to
\begin{equation}
F_{V}^2 \, M_{V}^2\, -\, F_{A}^2 \, M_{A}^2  \,=\, 0 \,  .
\label{eq:2ndWSR-LO}
\end{equation}
Therefore, if both WSRs are valid, $M_A > M_V$ and
the vector and axial-vector couplings are determined at LO in terms of the resonance masses:
\begin{equation}
\label{eq:FV_FA}
F_V^2\, =  \, v^2\, \Frac{M_A^2}{M_A^2-M_V^2} \, ,
\qquad \qquad \qquad F_A^2\, =\,v^2\, \frac{M_V^2}{M_A^2-M_V^2}\, .
\end{equation}

\subsubsection{WSR  constraints at one loop}

At high energy ($s\gg M_V^2, M_A^2, v^2$), the computed spectral functions (see Appendix \ref{app.spectralfunctions})
behave as:
\begin{eqnarray}\label{eq:Pipp}
\left.\mathrm{Im} \Pi_{30}(s)\right|_{\pi\pi} & = &
\frac{g^2 \tan{\theta_W}}{192\pi}\;\left\{\,
s \;\left(1-\frac{F_V G_V}{v^2}\right)^2\, +\; {\cal O}(s^0)\,\right\}\, ,
\\[10pt]\label{eq:PiVp}
\left.\mathrm{Im} \Pi_{30}(s)\right|_{V\pi} & = &
\frac{g^2 \tan{\theta_W}}{192\pi}\;\left\{\,
-\frac{s^2}{M_V^2 v^2} \;\left[F_V -2 G_V - F_A (2\kappa + \sigma)\right]^2
\, +\; {\cal O}(s)\,\right\}\, ,
\\[10pt]\label{eq:PiAp}
\left.\mathrm{Im} \Pi_{30}(s)\right|_{A\pi} & = &
\frac{g^2 \tan{\theta_W}}{192\pi}\;\left\{\,
\frac{s^2}{M_A^2 v^2} \;\left(F_A - F_V \sigma\right)^2\, +\; {\cal O}(s)\,\right\}\, .
\end{eqnarray}
Thus, their UV behaviour does not comply with the expected properties of the correlator
$\Pi_{30}(s)$. The two-particle spectral function Im$\Pi_{30}(s)$ must behave as a constant at high energies in order to make the dispersive integral~(\ref{eq.dispersivePi}) convergent.
Furthermore, the first WSR would demand that this constant term vanishes
and the second WSR would require the $1/s$ terms to be zero.

We will enforce that the sum of the three lowest-mass cuts, \ie the $\pi\pi$, $V\pi$ and $A\pi$
intermediate states, provides an acceptable representation of the correlator at short distances.
This means, that the sum of the three contributions should fall as $\cO(1/s)$
if at least the first WSR is assumed.
Imposing that the $\mO(s^2)$, $\mO(s)$ and $\mO(s^0)$ terms vanish provides three constraints
on the chiral couplings. It is important to highlight that
it is not possible to satisfy these constraints without the inclusion of the $\mathcal{L}_{VA}$
operators in Eq.~\eqref{VALagrangian} (the couplings $\kappa$ and $\sigma$), as it was already realized in Refs.~\cite{S-Isidori:08,S-Cata:10,S-Orgogozo:11}.
This result was known in QCD from the one-loop study of the pion form factor in $R\chi$T
\cite{L9a,L10,L9}.
We will also analyze the impact of imposing the second WSR as a fourth constraint; \ie
requiring the $\mO(1/s)$ term to also vanish.

After imposing the short-distance conditions on the spectral function,
one has to apply the same constraints to the real part of the correlator,
obtained through the dispersive calculation.
Using the high-energy expansion of the one-loop contribution,
\be
\overline{\Pi}(s) \; =\;
\Frac{ v^2}{s} \;\delta_{_{\rm NLO}}^{(1)} \, + \,
\Frac{v^2 \, M_V^2}{s^2}\;\delta_{_{\rm NLO}}^{(2)}\; +\; \cO\bigg(\Frac{1}{s^3}\bigg)\, ,
\ee
one reaches the NLO extension of the first and second WSRs~\cite{L8,L10,L8-Trnka},
respectively,
\begin{eqnarray}
F_{V}^{r\,2} \, -\, F_{A}^{r\,2}\; = \; v^2\, (1\,+\,\delta_{_{\rm NLO}}^{(1)}) \, ,\quad\;
\label{eq:NLO_WSR1} \\[10pt]
F_{V}^{r\,2}\, M_{V}^{r\,2} \, -\, F_{A}^{r\,2}\, M_{A}^{r\,2} \; = \;
v^2 \, M_{V}^{r\,2} \,\delta_{_{\rm NLO}}^{(2)}  \, .
\label{eq:NLO_WSR2}
\end{eqnarray}

If one assumes the two WSRs it is then possible to fix the couplings
up to NLO in the form,
\begin{eqnarray}
F_{V}^{r\,2}&=& v^2\; \frac{M_{A}^{r\,2}}{M_{A}^{r\,2}-M_{V}^{r\,2}}\;
        \left(1+\delta_{_{\rm NLO}}^{(1)}-\frac{M_{V}^{r\,2}}{M_{A}^{r\,2}}\,\delta_{_{\rm NLO}}^{(2)}  \right) \, ,
\label{FVr} \\[10pt]
F_{A}^{r\,2}&=& v^2\; \frac{M_{V}^{r\,2}}{M_{A}^{r\,2}-M_{V}^{r\,2}}\;
        \Bigl(1+\delta_{_{\rm NLO}}^{(1)}-\delta_{_{\rm NLO}}^{(2)} \Bigr) \, .
\label{FAr}
\end{eqnarray}
In the following,  we will use the renormalized masses $M_{R}^{r}$
in the NLO expressions and will denote them just as $M_R$.
The corrections have the structure
$\delta^{(k)}_{_{\rm NLO}}=\frac{M_V^2}{v^2} f^{(k)}(F_V,F_A,G_V,\kappa,\sigma, r)$,
and grow with the resonance mass ratio $r\equiv M_A/M_V$ as
$f^{(1)} \sim r^4$ and $f^{(2)}\sim r^6$, when $r\gg 1$.
Thus, a large mass splitting between the vector and axial-vector resonances
would lead to huge corrections over the LO result.

\subsection{Additional short-distance constraints}
\label{subsec:asdc}

Besides the conditions that come strictly from the analysis of the correlator $\Pi_{30}(s)$,
there are other constraints that have been considered in previous
works~\cite{S-Isidori:08,S-Cata:10,S-Orgogozo:11}:

\begin{itemize}
\item [--]{\bf $\mathbf{W_L W_L \rightarrow W_L W_L}$\ scattering}

The requirement that the tree-level $\pi\pi\to\pi\pi$
($W_L W_L\to W_L W_L$)  partial-wave scattering amplitudes behave like $\cO(s^0)$
at high energies leads to~\cite{S-Isidori:08,Bagger:1993zf}
\be\label{eq.scattering}
G_V\; =\; \Frac{v}{\sqrt{3}}\; .
\ee
This relation was already found in QCD at LO~\cite{juanjo-guo}
and at NLO~\cite{L9} in $1/N_C$, with $N_C=3$ the number of quark colours,
and assumes that the LO amplitude has a high-energy behaviour
similar (or as close as possible) to  that from the full scattering amplitude.
This might be a too strong constraint as quantum loops play a crucial role
in QCD and the perturbative relation \eqn{eq.scattering}
is not needed phenomenologically \cite{Nieves:2011gb}.\footnote{
Actually, the constraint (\ref{eq.scattering}) is not enough to satisfy the unitarity bound
at very high energies because the LO amplitude would still grow logarithmically as
$\ln{(s/M_V^2)}$.
This is a generic feature of the spin--1 meson exchanges in the crossed channels and
it is expected to be cured by higher-spin resonance exchanges.}

\item [--]{\bf  Vector form factor}

The two-Goldstone matrix element of the vector current defines the so-called
vector form factor (VFF),  $\langle \pi(p_1)\pi(p_2)|V^\mu|0\rangle = (p_1-p_2)^\mu F_\pi(s)$,
with $s = (p_1+p_2)^2$. At LO, it gets a direct constant contribution from the two-Goldstone
coupling to the vector current plus a vector-exchange term proportional to $F_V G_V/v^2$.
Imposing $F_\pi(s)$ to vanish at $s\rightarrow \infty$, one gets the LO constraint
\cite{RChT}
\be\label{VFF}
F_V G_V \; =\; v^2 \, .
\ee
This condition is equivalent to imposing a vanishing two-Goldstone spectral function at
short distances because
$\left.\mathrm{Im} \Pi_{30}(s) \right|_{\pi\pi} \sim s \; \left| F_\pi(s)\right|^2$
[see Eq.~(\ref{eq.SpecPP})].

\item [--]{\bf  Axial form factor}

The matrix element of the axial current between one Goldstone and one photon is parameterized
by the so-called axial form factor (AFF),
which at LO gets vector-exchange and axial-exchange contributions.
Requiring the AFF to vanish at $s\to\infty$ implies that
\be\label{AFF}
F_V - 2\, G_V \; =\; F_A\,\left(2\kappa +\sigma\right)\, .
\ee
In the absence of the $\mathcal{L}_{VA}$
operators in Eq.~\eqref{VALagrangian}, this would reduce to the well-known LO relation $F_V = 2\, G_V$ \cite{RChT}. The constraint \eqn{AFF} also guarantees that the leading $\mO(s^2)$ contribution to $\left.\mathrm{Im}\Pi_{30}(s)\right|_{V\pi}$
vanishes identically [see Eq.~\eqn{eq:PiVp}]. Since
$\mathrm{Im}\Pi_{30}(s)$ should not grow at large energies, we must then enforce the leading term in \eqn{eq:PiAp} to also vanish, which gives the additional condition
\be\label{ApSF}
F_A\; =\; F_V\,\sigma\, .
\ee

\end{itemize}

Combining the VFF constraint in Eq.~\eqn{VFF} with the AFF condition \eqn{AFF} and with Eq.~\eqn{ApSF}, one obtains (we choose the convention $F_V>0$ )
\be \label{eq.VFFAFF}
F_V\; =\; \frac{F_A}{\sigma}\; =\; \frac{v^2}{G_V}\; =\;
\frac{v}{\Omega(\kappa,\sigma)}\, ,
\qquad\qquad\qquad
\Omega(\kappa,\sigma) \;\equiv\;\sqrt{\frac{1-\sigma\, (2\kappa +\sigma)}{2}}\, ,
\ee
with $\sigma\, (2\kappa +\sigma) < 1$. If one also imposes the first WSR at LO,
all couplings get determined in terms of $\sigma$:
$
\kappa  = \frac{\sigma^2-1}{2\,\sigma}\, ,\;
\Omega = \sqrt{1-\sigma^2}\, ,\;
\sigma^2<1\, .
$
The second WSR, at LO, would then
require the ratio of resonance masses to take the value $M_A/M_V = 1/|\sigma|\, >\, 1$.
With\ $\sigma = -2\,\kappa = 1/\sqrt{2}$, one recovers the usual choice of LO
parameters \cite{RChT}:\
$F_V = \sqrt{2}\, v$, $G_V = v/\sqrt{2}$, $F_A = v$ \ and \ $M_A = \sqrt{2}\, M_V$.
The unitarity condition in Eq.~\eqn{eq.scattering} is obtained for\
$\Omega = 1/\sqrt{3}$; \ie $\sigma =\sqrt{2/3}$\ and\ $M_A/M_V =\sqrt{3/2}$.

\section{Phenomenology}
\label{sec.pheno}

The global fit to precision electroweak data provides the ``experimental'' value
\cite{Gfitter,LEPEWWG,ZFITTER}
\be\label{eq.Sexp}
S\; =\; 0.04\pm 0.10\, ,
\ee
normalized to the SM reference point $M_H = 0.120$ TeV.
We will also use $G_F = 11.663\, 788 \pm 0.000\, 007 \: \mathrm{TeV}^{-2}$
and $v = (\sqrt{2} G_F)^{-1/2}= 0.246\: \mbox{TeV}$ \cite{PDG}.

\subsection{LO results}

Let us first analyze the impact of the different short-distance contraints on the
LO prediction for the $S$ parameter in Eq.~\eqn{eq.S_LO}.

\begin{enumerate}
\item If one considers both the first and the second  WSRs, $F_V$ and $F_A$ take the
values in Eq.~(\ref{eq:FV_FA}), and $S_{\mathrm{LO}}$ becomes~\cite{Peskin:92}
\begin{equation}
S_{\mathrm{LO}}\; =\; \frac{4\pi v^2}{M_V^2}\,  \left( 1 + \frac{M_V^2}{M_A^2} \right) \, .
\label{eq.LO-S+2WSR}
\end{equation}
Since the WSRs imply $M_A>M_V$, the prediction
turns out to be bounded by
\begin{equation}
\frac{4\pi v^2}{M_V^2}\;\,\mathrm{max}\left(1\, ,\,\frac{2}{r^2}\right)
\; < \; S_{\rm   LO} \; < \;   \frac{8 \pi v^2}{M_V^2} \, , \label{SLOtwoWSR}
\end{equation}
with $r=M_A/M_V$. The additional VFF and AFF constraints in Section~\ref{subsec:asdc} lead to
$1/r^2 = \sigma^2< 1$. Imposing the unitarity condition~\eqn{eq.scattering},
$1/r^2 =2/3$\ and\ $S_{\mathrm{LO}}= 20\pi v^2/(3 M_V^2)$,
while the usually adopted choice\ $\sigma^2=1/2$\ gives\
$S_{\mathrm{LO}}= 6\pi v^2/M_V^2$.
\item If only the first WSR is considered,
and assuming  $M_A>M_V$,   one obtains for $S$ the lower bound
\begin{equation}
S_{\mathrm{LO}}
\; =\; 4\pi \left\{ \frac{v^2}{M_V^2} + F_A^2 \left( \frac{1}{M_V^2} - \frac{1}{M_A^2} \right)
\right\}\; >\; \frac{4\pi v^2}{M_V^2}  \; >\; \frac{4\pi v^2}{M_A^2}    \, .
\label{eq.LO-S+1WSR}
\end{equation}
Thus, $S_{\mathrm{LO}}$ is predicted to be positive, provided $M_A>M_V$.
The VFF and AFF conditions determine $F_A^2/v^2 = \sigma^2/(1-\sigma^2)$. Therefore,
$\sigma^2=2/3$ would imply\
$S_{\mathrm{LO}}= 12\pi v^2\, [1 - 2/(3r^2)]/M_V^2$, while\ $\sigma^2=1/2$\ gives
$S_{\mathrm{LO}}= 8\pi v^2\, [1 - 1/(2r^2)]/M_V^2$.

The possibility of an inverted mass ordering of the vector and axial-vector resonances in vector-like SU(N) gauge theories, close to a conformal transition region, was considered in \cite{Appelquist:1998xf}. The LO lower bound on $S_{LO}$ becomes then an upper bound:   $S_{\mathrm{LO}} < 4\pi v^2/M_V^2 < 4\pi v^2/M_A^2 $. Note that if the splitting of the vector and axial-vector resonances was small, the prediction of $S_{LO}$ would be close to the upper bound and the main conclusion of this section would be stable.
\end{enumerate}

\begin{figure}[t]
\begin{center}
\includegraphics[scale=1,width=7.75cm]{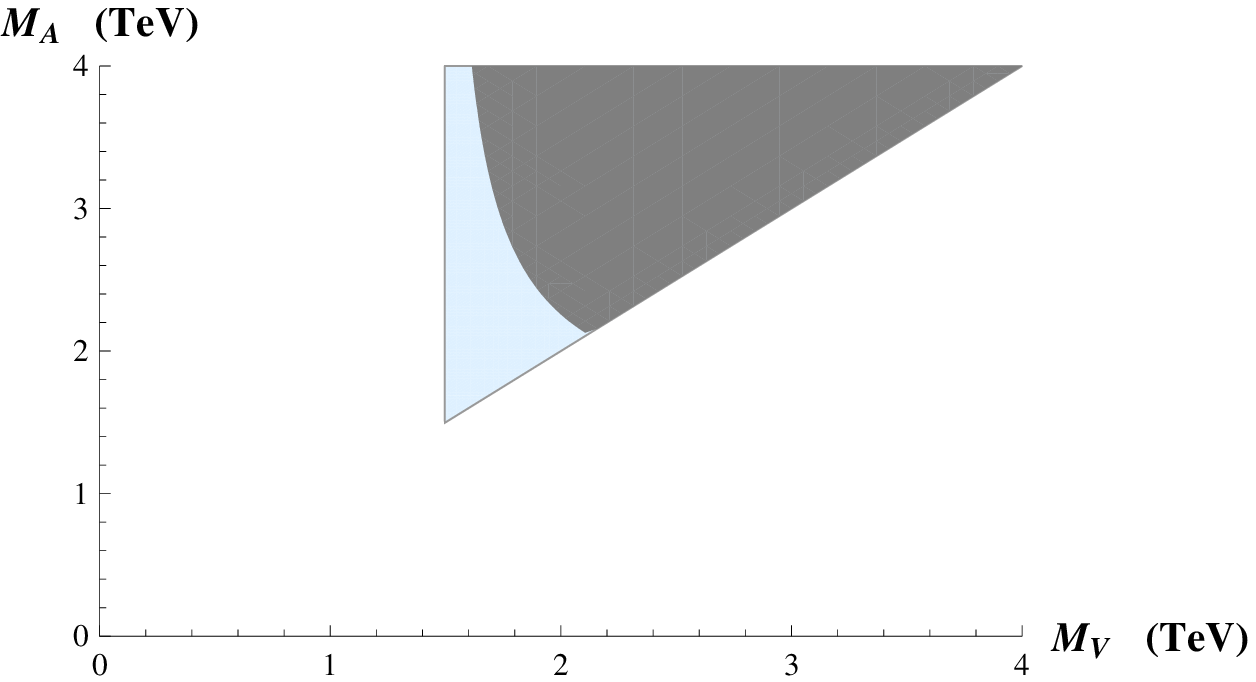}
\hskip .5cm
\includegraphics[scale=1,width=7.75cm]{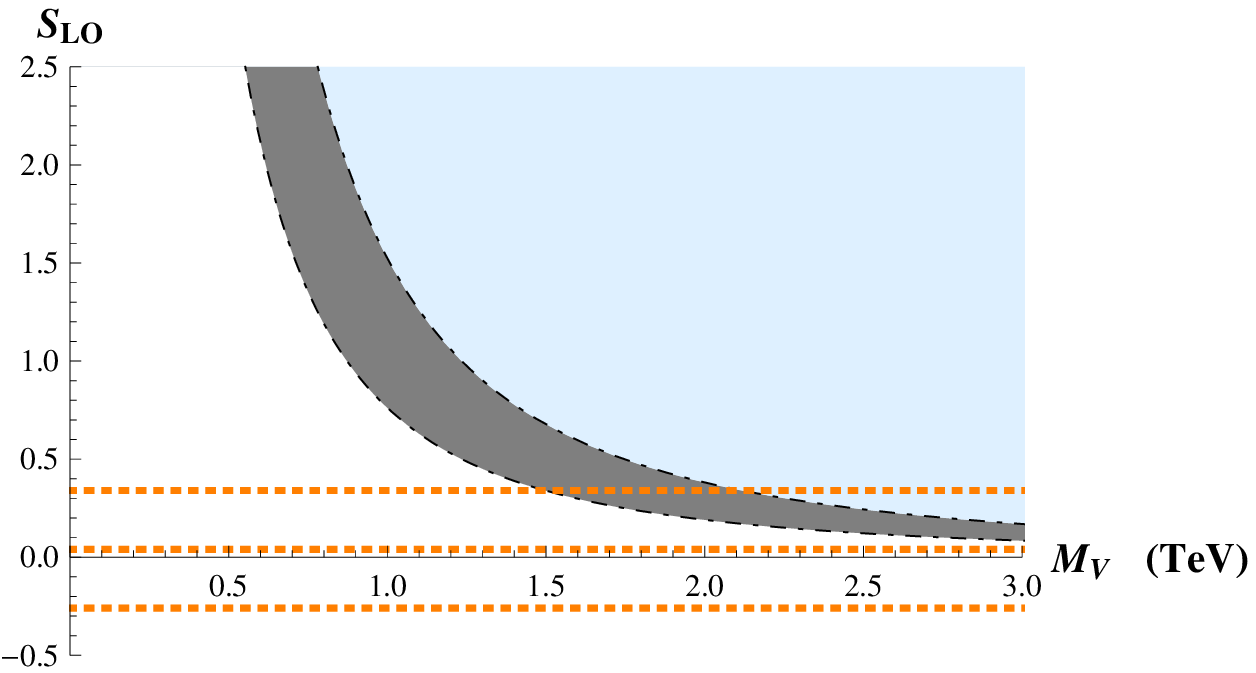}
\caption{\small
Regions for $M_V$ and $M_A$ where $S_{\mathrm{LO}}$ is compatible with the
data at the $3\sigma$ level (left) and LO predictions for $S$ (right).
The dark gray regions
correspond to Eqs.~(\ref{eq.LO-S+2WSR}) --left--
and (\ref{SLOtwoWSR}) --right--, which
take into account the two WSRs. The lower bound~(\ref{eq.LO-S+1WSR}),
which only assumes the first WSR and $M_A>M_V$,
is satisfied in the light-blue regions in addition to the dark-gray ones.
The horizontal dotted lines on the right correspond to the experimentally allowed region at $3\sigma$.}
\label{fig.allowed-LO}
\end{center}
\end{figure}

The resonance masses need to be heavy enough to comply with the strong experimental
bound in Eq.~\eqn{eq.Sexp}. In Figure~\ref{fig.allowed-LO} we show the ranges of
resonance masses, $M_V$ and $M_A$, which are compatible
with the experimental data at the $3\sigma$ level.
The dark gray region assumes the two WSRs,
while the allowed range gets enlarged to the light-blue region if the second WSR is relaxed
and one only assumes the first WSR and $M_A> M_V$.
Even with the softer requirements, the experimental data implies $M_V>1.5$~TeV (2.3~TeV) at the
3$\sigma$ (1$\sigma$) level.
The figure on the right compares the corresponding LO predictions with the experimentally
allowed region at 3$\sigma$.

\subsection{NLO results imposing both Weinberg sum rules}

\subsubsection{Imposing both WSRs at LO and NLO}
\label{sec.2WSR-constraintsA}

At LO, in order to fulfill the two WSRs, one needs to consider the exchanges
of at least one vector and one axial states (in addition to the Goldstone pole).
Similarly, at NLO the minimum number of
two-particle cuts needed to satisfy both WSRs are the $\pi\pi$, $V\pi$ and $A\pi$
intermediate states. We are assuming that these contributions dominate
the dispersive integral~(\ref{eq.S-dispersive}).

\begin{table}[t!]
\begin{center}
\begin{tabular}{|c |c |c ||c|c|c|c|}
\hline   &&&&&&
\\[-9pt]
$\;  \Frac{M_A}{M_V} \; $  &  $\;\sigma\;$   &  $\;\kappa\;$
&   $1- \Frac{F_V G_V}{v^2}  $   &  $1-\Frac{3 G_V^2}{v^2} $   &
$  \Frac{F_V\! -\! 2 G_V\! -\! F_A(2 \kappa\! +\!\sigma) }{v} $ &
$  \Frac{F_A\! -\! F_V\sigma}{v} $
\\[7pt]
\hline  &&&&&&
\\[-9pt]
1.02 &    1.03 &    -0.05 &    0.49 &    0.97 &    0.25 &    -0.25    \\
1.08 &    -0.03 &    0.90 &    0.16 &    0.68 &    -2.26 &    2.45    \\
1.55 &    -3.63 &    2.49 &    -1.46 &    -9.63 &    -3.59 &    5.58    \\
\bf{1.58} &   \bf{ 0.65} &  \bf{  -0.34} & \bf{   0.15} &  \bf{  -0.30} &  \bf{  0.01} &  \bf{  -0.02}    \\
\bf{1.69} &  \bf{  0.60} &  \bf{  -0.30} &  \bf{  0.22} &  \bf{  -0.18} &  \bf{  -0.01} &  \bf{  -0.02}    \\
1.75 &    0.40 &    0.35 &    0.65 &    0.75 &    -0.12 &    0.21    \\
1.94 &    0.12 &    0.30 &    0.43 &    0.28 &    -0.24 &    0.46    \\
3.09 &    0.24 &    0.33 &    0.62 &    0.61 &    0.03 &    0.09    \\
[2pt]
\hline
\end{tabular}
\caption{\small
Set of eight real solutions to the two WSRs, considered both at the LO
and NLO
($\pi\pi$, $V\pi$ and $A\pi$ cuts).
The last four  columns indicate how well they fulfill
the additional high-energy constraints in Section~\ref{subsec:asdc}.
The most reasonable solutions are indicated in boldface.
}
\label{tab.1+2WSR-sols}
\end{center}
\end{table}

The requirement that the NLO correlator $\Pi_{30}(s)$ vanishes like $1/s^2$
at short distances determines  the ``renormalized'' resonance couplings
through Eqs.~(\ref{eq:NLO_WSR1})
and (\ref{eq:NLO_WSR2}).
Using Eq.~\eqn{eq.Sbar}, this leads to the low-energy prediction
\bear
S_{\rm NLO}
& =& \, 4\pi v^2 \, \bigg[\Frac{1}{M_{V}^{r\, 2}} +\Frac{1}{M_{A}^{r\, 2}}\bigg]
\;  \bigg(1+\delta_{\rm NLO}^{(1)}
-  \frac{M_{V}^{r\, 2} \delta_{\rm NLO}^{(2)}}{M_{V}^{r\, 2}+M_{A}^{r\, 2}} \bigg) \, \; +\,\;\overline{S}\, .
\eear

The four short-distance constraints on the two-particle spectral function
(absence of $\mO(s^2)$, $\mO(s)$, $\mO(s^0)$ and $\mO(1/s)$ terms, {\it i.e.}, $c_{-2}=c_{-1}=c_0=c_1=0$ in Appendix B) plus the two
NLO (LO) WSRs allow us to determine 6 parameters: $F_V^r$
($F_V$), $F_A^r$ ($F_A$), $G_V$, $\kappa$, $\sigma$ and the mass ratio $M_A^r/M_V^r$ ($M_A/M_V$).
We discard solutions with $G_V <0$
because they strongly violate the VFF condition \eqn{VFF}.
We found eight real sets of solutions satisfying all constraints,
which are listed in Table~\ref{tab.1+2WSR-sols}.
The table also shows how well each solution satisfies the additional constraints in Section~\ref{subsec:asdc} (VFF, unitarity, AFF).
The solution with $r=M_A/M_V=1.55$ can be clearly discarded, as it grossly violates
all short-distance conditions. The one with $r=1.02$ does not look too good either;
it implies a large departure of the unitarity condition \eqn{eq.scattering}.
Notice that these two bad solutions have $\sigma^2>1$.
The best solutions are the ones with $r=1.58$ and $r=1.69$ which satisfy all
constraints within better than 30\%.

\begin{figure}[t!]
\begin{center}
\includegraphics[scale=1,width=7.75cm]{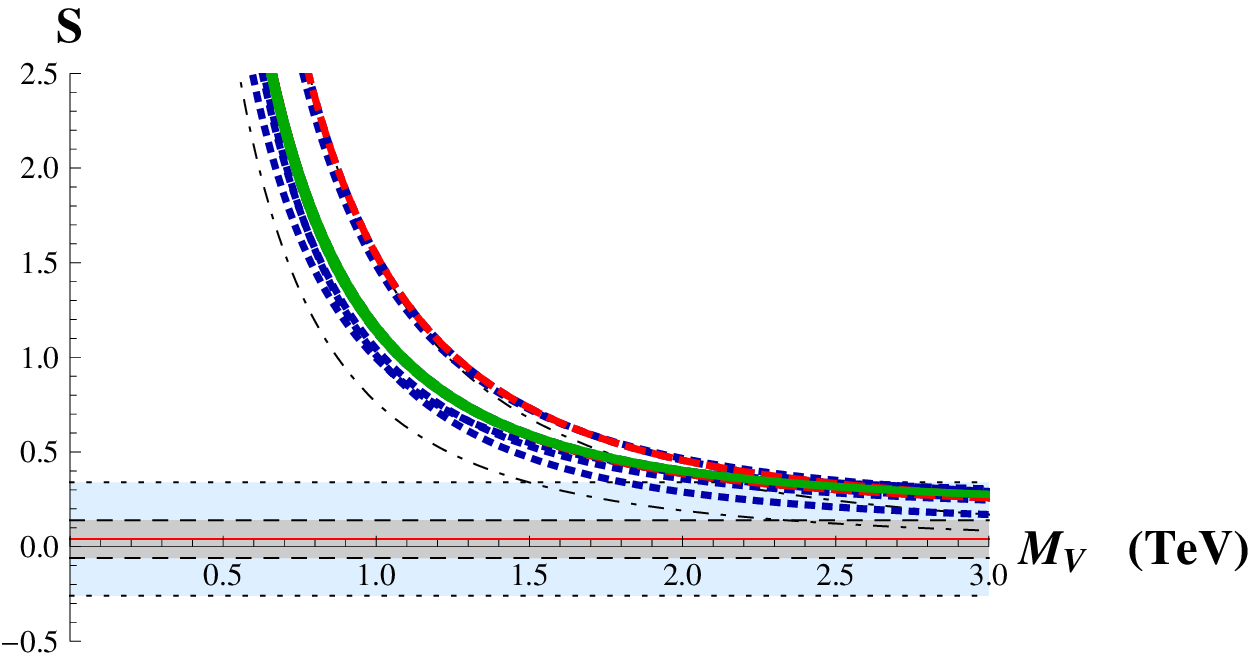}
\hskip .5cm
\includegraphics[scale=1,width=7.75cm]{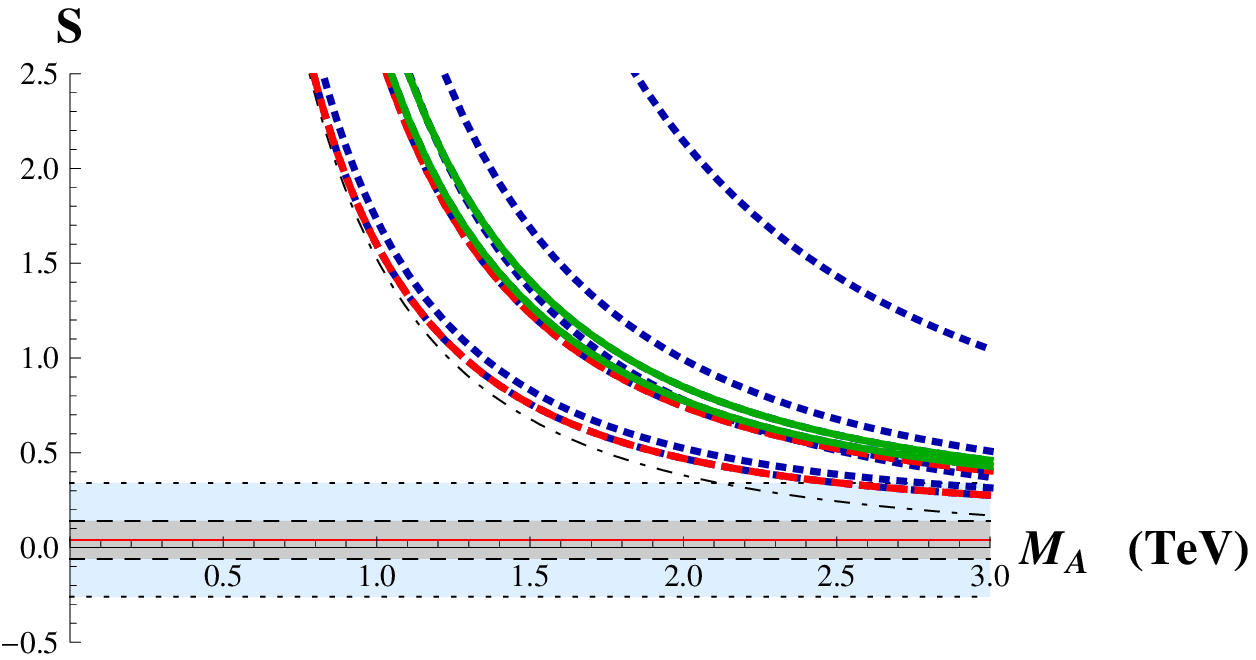}
\caption{\small NLO determination of $S$, imposing the two WSRs.
The 2 optimal solutions correspond to the continuous green curves.
The dash-dotted curves  provide the LO bounds
from Eq.~(\ref{SLOtwoWSR}).
The horizontal dashed (dotted)
lines show the experimentally allowed region at $1\sigma$ ($3\sigma$).
The red horizontal line is the experimental central value.
}
\label{justWSR+LO-S}
\label{fig.MaS_1+2WSR}
\end{center}
\end{figure}

The corresponding predictions for $S_{\mathrm{NLO}}$ are shown in Figure~\ref{justWSR+LO-S}
as a function of the vector (left) or axial-vector (right) masses.
The continuous green curves indicate the two optimal solutions,
while the 6 additional possibilities are given by the dotted blue curves, with the exception of the 2 bad solutions $r=1.02$ and $r=1.55$, which are plotted in dashed red.
The dash-dotted curves  provide the LO bounds
$\frac{4\pi v^2}{M_V^2}<S_{LO}< \frac{8\pi v^2}{M_V^2}$
and  $\frac{8\pi v^2}{M_A^2} < S_{\rm LO}$  from Eq.~(\ref{SLOtwoWSR}).

The smooth UV behaviour of the spectral function implies a well-behaved one-loop contribution.
Therefore, the differences with respect to the LO estimate are not very large.
In order to obtain a  value of $S$ compatible with the experimental band,
one needs roughly the same range of masses as at tree-level.
At NLO, we find that $M_V>1.8$~TeV (3.8~TeV) at the 3$\sigma$ (1$\sigma$) level.
The resulting bounds on the axial mass are much stronger, requiring $M_A > 2.5$~TeV  (6.6~TeV) at 3$\sigma$ (1$\sigma$).
We can safely conclude that in Higgsless electroweak theories satisfying the two WSRs, such as the usual technicolour models, the associated
spectrum of vector and axial-vector resonances should be heavier than the 1 TeV scale.

\subsubsection{Imposing both WSRs at NLO and the VFF and AFF constraints}
\label{sec.2WSR-constraintsB}

In the previous analysis we have used the LO WSRs to fix the Lagrangian parameters
$F_V$ and $F_A$, as indicated in Eq.~\eqn{eq:FV_FA}. The values of these two decay constants
are slightly shifted at the NLO, but their LO expressions are good enough to parameterize
the spectral function Im$\Pi_{30}(s)$. Any difference
between the LO and NLO decay constants amounts to a higher-order effect, which can be neglected
in the NLO evaluation of $S$. Nevertheless, it is worth investigating the possible size
of these corrections.

An alternative procedure to fix the $\Pi_{30}(s)$ correlator at the NLO is to use the
VFF and AFF constraints in Section \ref{subsec:asdc}.
Thus, we will drop now the LO WSRs~(\ref{eq:1stWSR-LO}) and (\ref{eq:2ndWSR-LO})
and use instead the relations~(\ref{VFF}) and (\ref{AFF}).
The WSRs will only be applied at the NLO, imposing the
short-distance requirements for the two--particle spectral function Im$\Pi_{30}(s)$
and the full NLO correlator [Eqs. \eqn{FVr} and \eqn{FAr}].
We will check {\it a posteriori}  how well the LO WSRs are obeyed by our solutions.

Imposing the two WSRs at the NLO implies that $\mathrm{Im}\Pi_{30}(s)$
behaves like  $\mathcal{O}(1/s^2)$; {\it i.e.}, $c_{-2}=c_{-1}=c_0=c_1=0$ \ in Appendix~B.
We have then 7 parameters and 6 constraints, which allows us to fix
$F_V$, $F_A$, $G_V$, $\kappa$, $\sigma$ and the mass ratio $r=M_A/M_V$.
We will leave the vector mass $M_V$ as a free parameter.
We follow five steps to extract the valid solutions:
\begin{enumerate}
\item
Considering $c_{-2}=0$, the VFF and AFF constraints,
and choosing the convention $F_V>0$,
one obtains the relation \eqn{eq.VFFAFF}.
\item
Setting   $c_{-1}=0$, we  find
\begin{equation}
\kappa \,=\, \pm \frac{1-\sigma^2}{2} \equiv \kappa_\pm \,.
\end{equation}
It can be easily checked that $(\kappa_+, \sigma)$ and $(\kappa_-, -\sigma)$
are equivalent solutions because both lead to
the same spectral function $\mathrm{Im}\Pi_{30}(s)$  and the same value for $\Omega$,
\begin{equation}
\Omega(\kappa_+, \sigma) \,=\, \Omega(k_-, -\sigma) \,  =  \, \sqrt{\frac{1-\sigma-\sigma^2+\sigma^3}{2}}\,.
\end{equation}
Therefore we will only keep one of them.
\item
Imposing $c_{0}=0$ we find four non-equivalent real solutions for $\sigma$ in terms of $r=M_A/M_V$.
However, we discard two of them because they
give $(F_V^2-F_A^2)/v^2 <0$,  strongly violating the first WSR at LO.
\item
The second WSR at NLO,
{\it i.e.}, $c_{1}=0$, finally determines $r$ for each of the previous two solutions.
\item
In the last step, we determine the ``renormalized'' resonance couplings
$F_{V}^{r}$ and $F_{A}^{r}$  through the NLO relations~(\ref{FVr})
and~(\ref{FAr}).
\end{enumerate}

\begin{table}[t!]
\begin{center}
\begin{tabular}{|c |c |c ||c|c|c|}
\hline   &&&&&
\\[-9pt]
$\;  \Frac{M_A}{M_V} \; $  &  $\;\sigma\;$   &  $\;\kappa\;$
&     $\;1-\Frac{3 G_V^2}{v^2}\; $   & $\;(F_V^2-F_A^2-v^2)/v^2\;$ & $\;(F_V^2 M_V^2 -F_A^2 M_A^2)/(v^2 M_V^2)\;$
\\[7pt]
\hline  &&&&&
\\[-9pt]
1.01    &    1.00    &    0.00    &    1.00    &    0.00    &    -451.45    \\
{\bf 1.65 }   &  {\bf  0.61 }    &  {\bf  -0.32}    &  {\bf  -0.53}    &  {\bf  0.25 }   &  {\bf  0.01}    \\
[2pt]
\hline
\end{tabular}
\caption{\small
Solutions to the two WSRs, at the NLO, and the VFF and AFF short-distance contraints.
The last two  columns indicate how well they fulfill
the two LO WSRs~(\ref{eq:1stWSR-LO}) and (\ref{eq:2ndWSR-LO}).
The only acceptable solution is indicated in boldface.
}
\label{tab.WSRsLO}
\end{center}
\end{table}

\begin{figure}[t!]
\begin{center}
\includegraphics[scale=1,width=7.75cm]{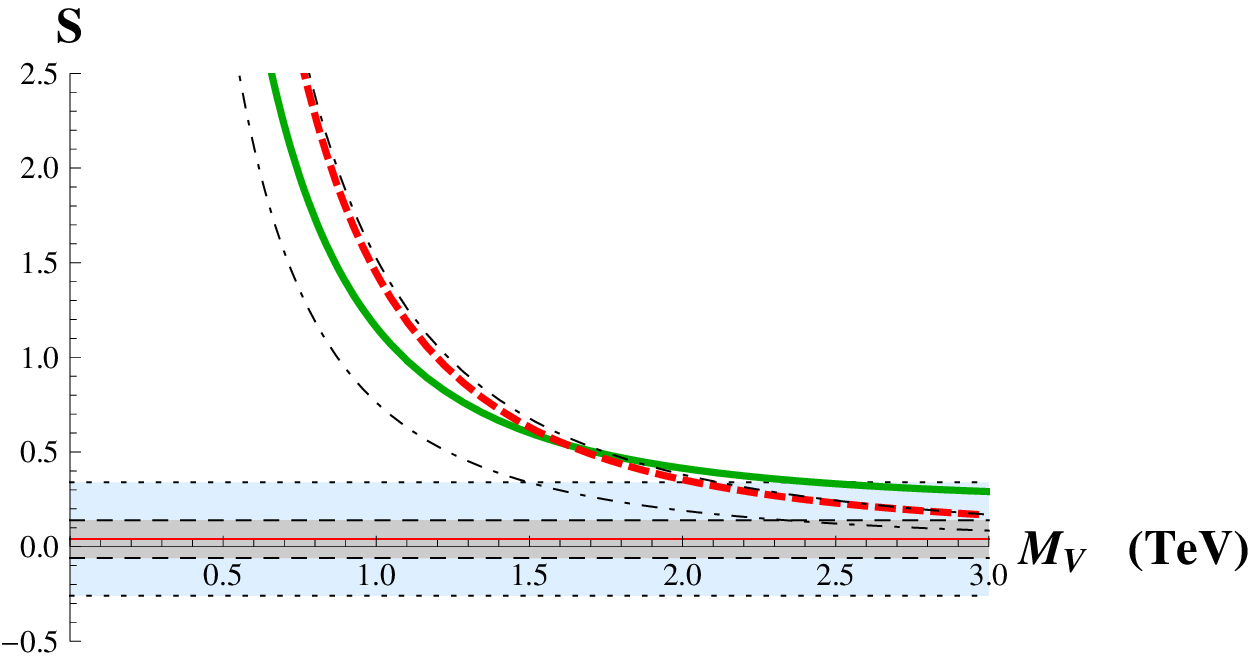}
\hskip .5cm
\includegraphics[scale=1,width=7.75cm]{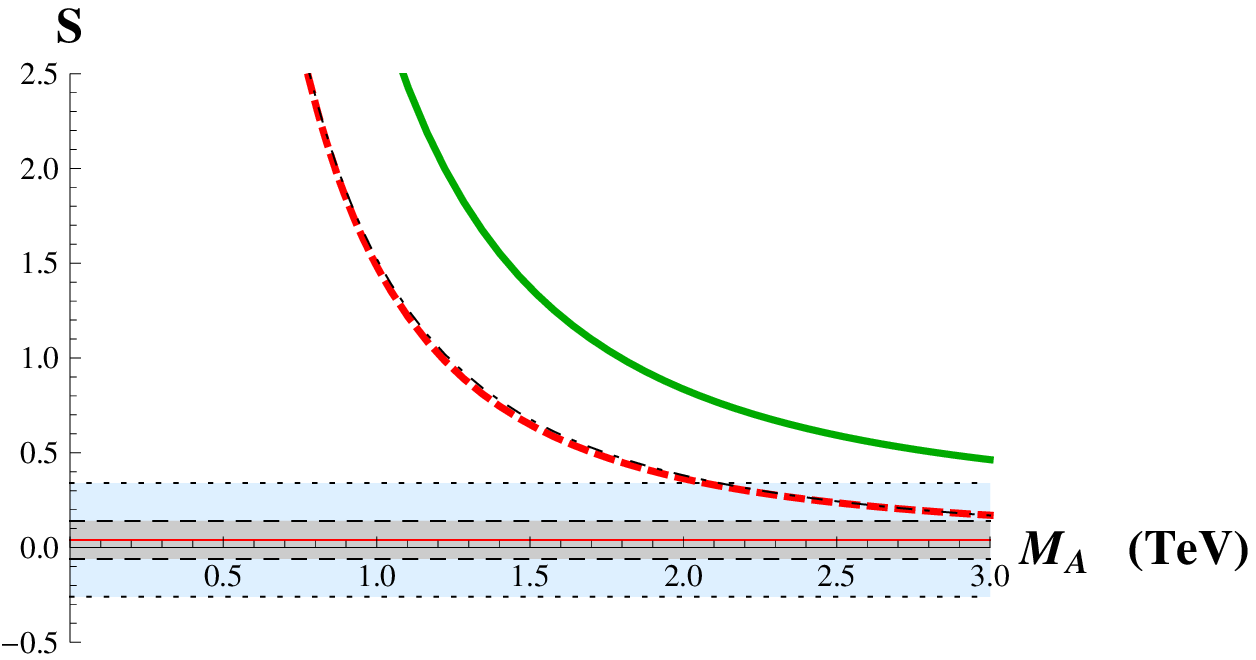}
\caption{\small NLO determination of $S$, as a function of $M_V$ and $M_A$,
imposing the two WSRs plus the VFF and AFF constraints.
The solid green and dashed red curves refer to the {\it good} and {\it bad} solutions.
The dash-dotted curves  provide the LO bounds
from Eq.~(\ref{SLOtwoWSR}).
Same horizontal experimental bands as in Figure~\ref{justWSR+LO-S}.
}
\label{fig.2WSR+VFF+AFF}
\end{center}
\end{figure}

In Table~\ref{tab.WSRsLO},  we list the two solutions and
study how well the LO WSRs~(\ref{eq:1stWSR-LO}) and (\ref{eq:2ndWSR-LO})
are satisfied.  The solution with $r=1.01$
corresponds to $1-\sigma=3\cdot  10^{-5}$ and
$\kappa=- 3\cdot 10^{-5}$, which have been rounded in the table
in order to ease the comparison; it
can be clearly discarded,
because it sharply violates the second WSR at LO.
This {\it bad} solution also violates the unitarity relation \eqn{eq.scattering}
and yields $F_{V}^{r\, 2},\, F_{A}^{r\, 2}<0$ for $M_V>1$~TeV.
Nonetheless, we have plotted  both the {\it good} ($r=1.65$) and {\it bad} ($r=1.01$) predictions
for $S$   in Figure~\ref{fig.2WSR+VFF+AFF},   as a function of $M_V$ and $M_A$,
because it will be helpful to better understand the next  analysis
with  just the first WSR.
Comparing with the results in Table~\ref{tab.1+2WSR-sols}, it is interesting to remark that
the {\it bad} solution is close to the solution
obtained in the  previous NLO analysis with $r=1.02$, while
the {\it good} solution is close to the optimal solutions with $r=1.58$ and $1.69$
found previously.

%

The comparison of the outcome from the {\it good} solution with the ``experimental'' data
sets a lower bound $M_V>2.4$~TeV ($M_A>4.0$~TeV) at the $3\sigma$ level. The closest agreement appears for $M_V\simeq 6.2$~TeV ($M_A\simeq 10.2$~TeV) at $2.0\sigma$.

\subsection{NLO results with just the first Weinberg sum rule}

While the second WSR is only expected to apply in QCD-like electroweak models,
the first WSR remains valid for a wider class of theories, including conformal
models~\cite{Peskin:92,S-Orgogozo:11}.
Therefore, it is relevant to analyze how our previous NLO predictions
vary when we relax the short-distance constraints,
dropping the second WSR and keeping just the first one.

Without the second WSR we can no-longer determine $F_V^r$ and $F_A^r$
[see Eqs.~(\ref{FVr}) and (\ref{FAr})]. Therefore, we can only derive lower bounds on $S$.
Using the first WSR relation~\eqn{eq:NLO_WSR1} in Eq.~\eqn{eq.Sbar},
and  assuming $M_A^r > M_V^r$, we obtain the inequality:
\begin{equation}
S_{\mathrm{NLO}}
\,>\, 4\pi \,v^2\,\mathrm{max} \left(
\frac{1+\delta_{_{\rm NLO}}^{(1)}}{M_V^{r\,2}},\frac{1+\delta_{_{\rm NLO}}^{(1)}}{M_A^{r\,2}}\right)\, +\, \overline{S}\, ,
 \label{lower-bound}
\end{equation}
which at LO reduces to Eq.~\eqn{eq.LO-S+1WSR}. As it has happened in the LO case, if we consider an inverted hierarchy of the vector and axial-vector resonances~\cite{Appelquist:1998xf}, $M_A < M_V$, the lower bound becomes an upper bound: $S_{NLO}  < 4\pi v^2\, \mathrm{min} \left( (1+\delta_{\mathrm{NLO}}^{(1)})/M_V^{r\,2} ,  (1+\delta_{\mathrm{NLO}}^{(1)})/M_A^{r\,2} \right)$. A small splitting between the resonances would imply a value of $S_{NLO}$ close to the upper bound and the conclusions of this work would not change appreciably.

We discuss next the results we obtain under different hypotheses.

\subsubsection{$\pi\pi$ channel only}

\begin{figure}
\begin{center}
\includegraphics[scale=1,width=11cm]{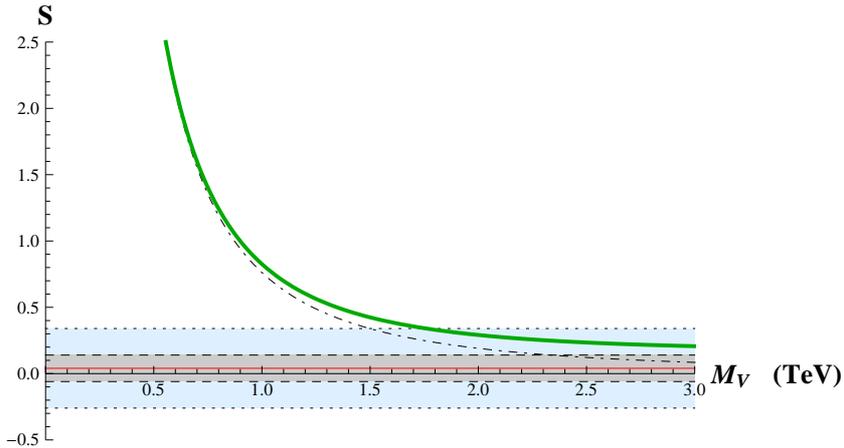}
\caption{\small
Lower bound on $S_{\mathrm{NLO}}$ as a function of $M_V$, including only
the $\pi\pi$ channel and assuming the first WSR (continuous curve).
The dash-dotted brown line is the LO bound $S>4\pi v^2/M_V^2$.
Same horizontal experimental bands as in Figure~\ref{justWSR+LO-S}.}
\label{1WSR+pipi}
\end{center}
\end{figure}

On the contrary to what happened with two WSRs, now it is possible to perform the analysis with only the lightest two-particle absorptive channel, {\it i.e.}, just the $\pi\pi$ cut.
The corresponding spectral function, given in Eq.~\eqn{eq.SpecPP}, grows linearly with $s$ unless
the relation $F_V G_V=v^2$ is satisfied. Thus, one recovers the VFF constraint in Eq.~(\ref{VFF}),
which in this case guarantees that $\left.\mathrm{Im}\Pi_{30}(s)\right|_{\pi\pi}\sim 1/s$.
This leads to the result
\be\label{eq.Spipi}
\bar S\, =\, \frac{1}{12\pi}\,\left[\ln{\left(\frac{M_V^2}{M_H^2}\right)}-\frac{17}{6}\right]\, ,
\qquad\qquad\qquad
\delta_{_{\rm NLO}}^{(1)}\, =\, \frac{M_V^2}{48\pi^2 v^2}\, ,
\ee
which only depends on the vector resonance mass and the chosen reference value of the Higgs mass, $M_H = 0.120$ TeV. Figure~\ref{1WSR+pipi} shows the resulting lower bound~(\ref{lower-bound})
as a function of $M_V$. It is similar to the LO bound, but slightly stronger, increasing the
tension with the data which now requires $M_V > 1.8$ TeV at the 3$\sigma$ level. The lower bound stays above the 1$\sigma$ experimental band for any value of $M_V$ (one reaches the closest agreement for $M_V\simeq 5.4$~TeV at the $1.4\sigma$ level).

\subsubsection{$\pi\pi$, $V\pi$ and $A\pi$ channels:
1st  WSR at NLO,  VFF and AFF constraints}

\begin{figure}
\begin{center}
\includegraphics[scale=1,width=5.1cm]{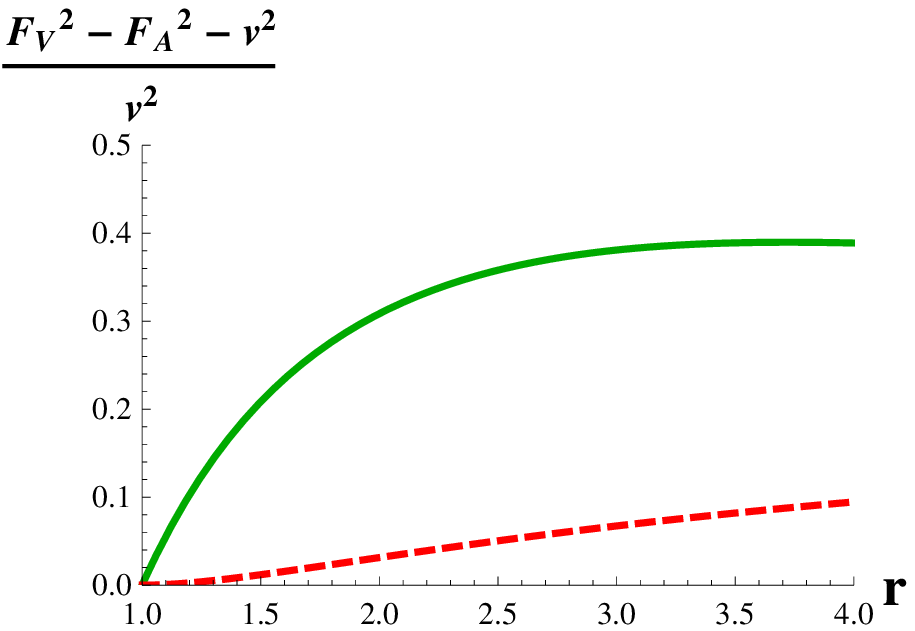}
\hskip .5cm
\includegraphics[scale=1,width=5.1cm]{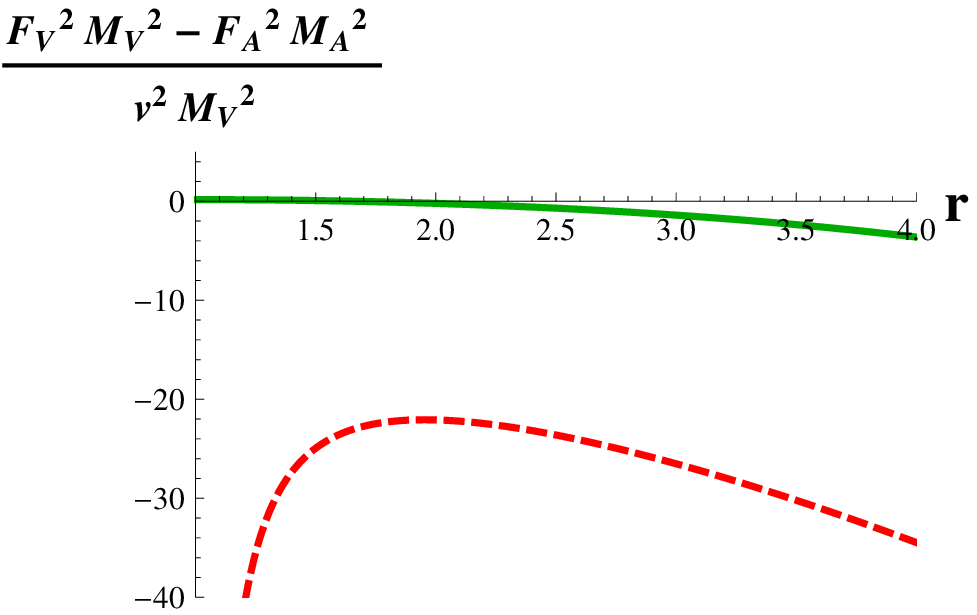}
\hskip .5cm
\includegraphics[scale=1,width=5.1cm]{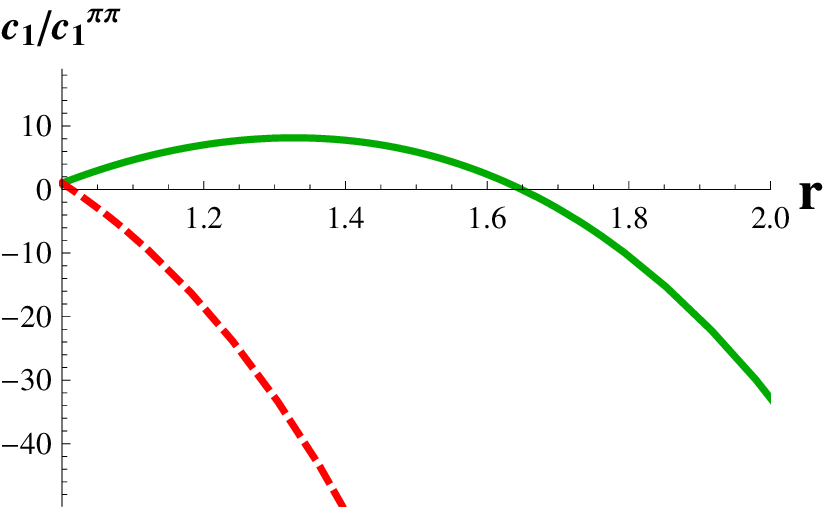}
\caption{\small Violation of the first (LO) and second WSRs (LO and spectral function at NLO)} for
the {\it good} (solid green) and {\it bad} (dashed red) solutions, being $c_1^{\pi\pi}=   v^4 M_A^2 M_V^6$.
\label{fig.checks-WSR}
\end{center}
\end{figure}

Following the same procedure adopted in Sec.~\ref{sec.2WSR-constraintsB}
but, this time, without imposing the second WSR,
we find two solutions  with $(F_V^2-F_A^2)/v^2>0$, where $r=M_A/M_V$
is left unfixed.  In Figure~\ref{fig.checks-WSR},   we have plotted
the predicted violation of
the two LO WSRs, Eqs.~(\ref{eq:1stWSR-LO})  and (\ref{eq:2ndWSR-LO}), and the NLO 2nd WSR (spectral function), Eq.(\ref{c1}),
for each of these two solutions,
as a function of $r$.  One of the solutions, denoted as {\it good}, obeys moderately well
both sum rules. On the other hand,  the other solution ({\it bad})
badly violates the second sum rule for any $r$, with
$(F_A^2 M_A^2-F_V^2M_V^2)/(v^2M_V^2)\gg 1$.

\begin{figure}
\begin{center}
\includegraphics[scale=1,width=7.75cm]{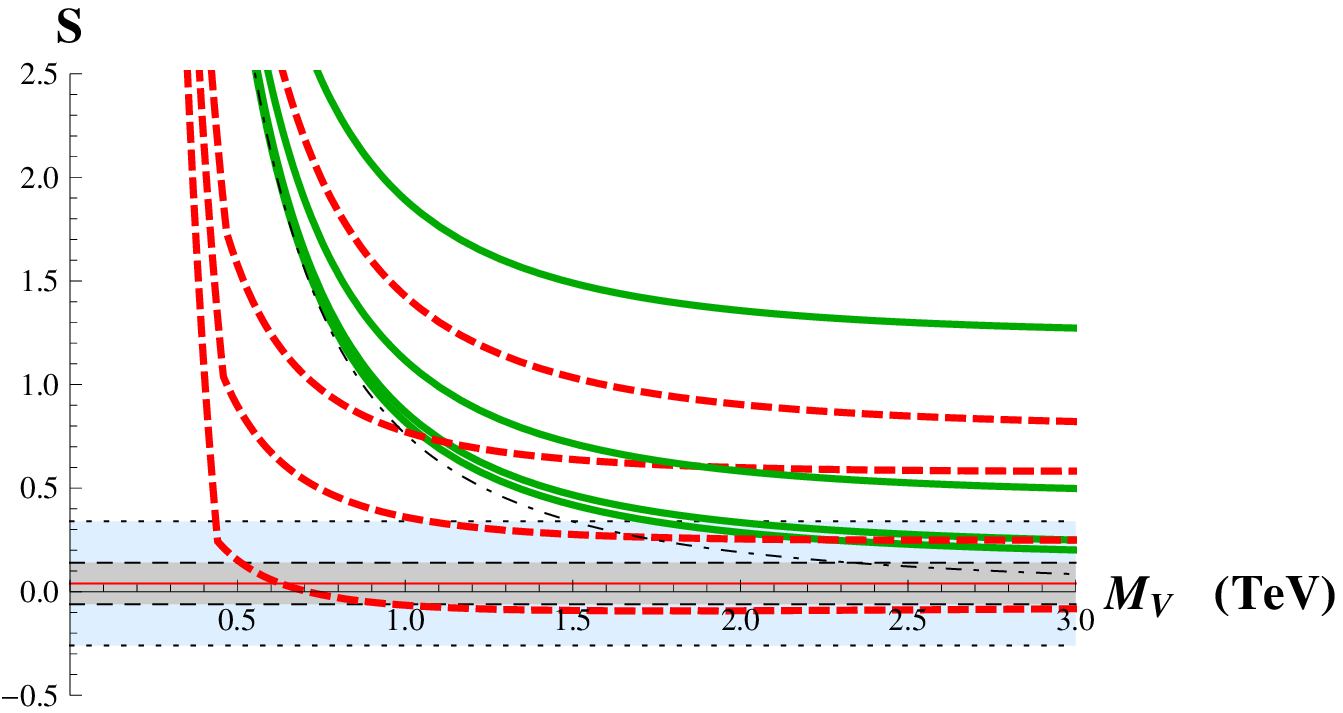}
\hskip .5cm
\includegraphics[scale=1,width=7.75cm]{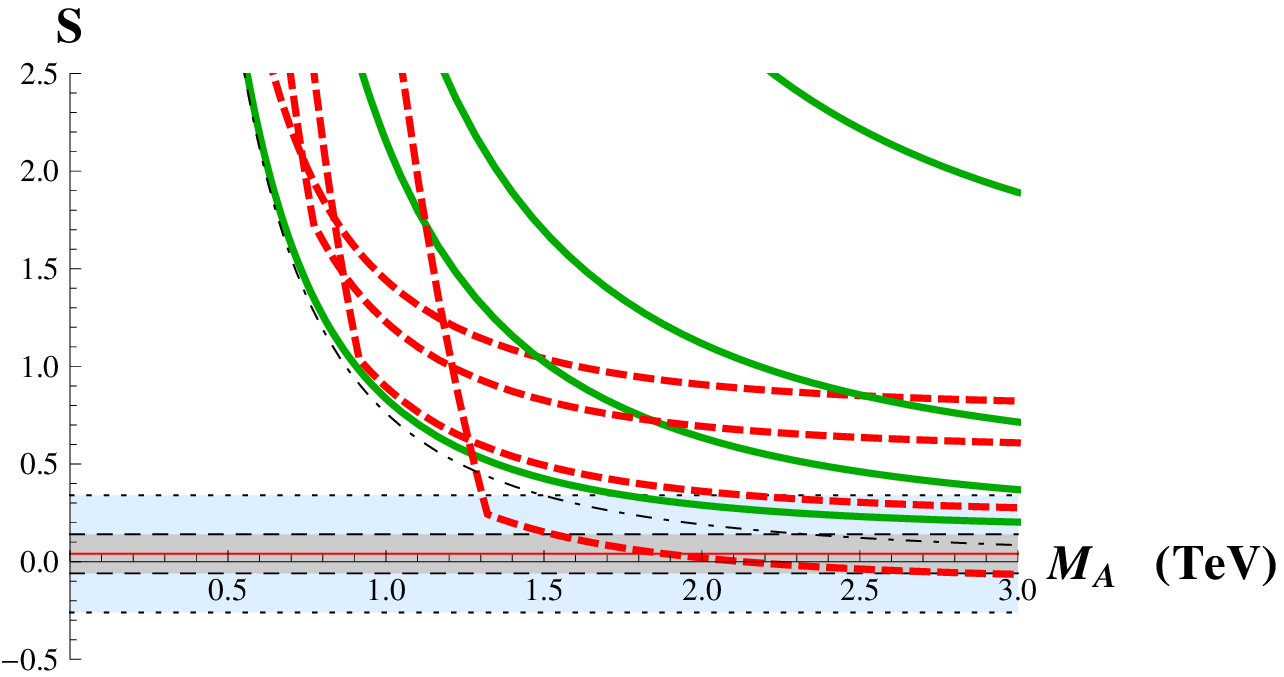}
\caption{\small
Lower bound on $S_{\mathrm{NLO}}$ as a function of
$M_V$ and $M_A$, imposing the first WSR plus the VFF and AFF constraints.
The green (continuous) and red (dashed) curves refer to the {\it good} and {\it bad} solutions.
We have considered the inputs $r=1.01,\, 1.65,\, 2.00,\, 3.00$.
As $r$ grows,    for the {\it good} ({\it  bad}) solution
the asymptotic value of $S$ at large masses  increases (decreases).
The dash-dotted curves  provide the LO bounds
$S_{LO}> 4\pi v^2/M_V^2$ and $S_{LO}> 4\pi v^2/M_A^2$ from
Eq.~(\ref{eq.LO-S+1WSR}).
Same horizontal experimental bands as in Figure~\ref{justWSR+LO-S}.
}
\label{fig.1WSR+VFF+AFF}
\end{center}
\end{figure}

Figure~\ref{fig.1WSR+VFF+AFF} shows the predicted lower bounds on $S$ for the two sets of solutions and various mass ratios: $r=1.01,\, 1.65,\, 2.00,\, 3.00$.
The first two values of $r$ correspond to the solutions found in Table~\ref{tab.WSRsLO},
imposing the two WSRs.

The {\it good\/} solution leads to lower bounds on $S$ which increase with increasing
values of the mass ratio $r$. Therefore, the absolute lower bound
is obtained for $r\to 1$.
One needs $M_V>1.8$~TeV ($M_A>1.8$~TeV)
to reach compatibility with  the ``experimental'' data at the $3\sigma$ level.
The closest agreement occurs for $M_V\simeq   5.4$~TeV, at the
$1.3\sigma$  level.

%

Conversely, in the case of   the {\it bad\/} solution,
the predicted lower bound on $S$, in terms of $M_V$,  decreases  when the mass splitting
between the vector and axial-vector states grows.    Thus,  the resulting lower bound for the vector mass decreases
when  $r$ increases and, for  $r>2.0$ ($r>2.3$),   one may find solutions for
$M_V$ below 1~TeV   at the $3\sigma$ ($1\sigma$) level.
However, the   $3\sigma$ ($1\sigma$) level lower bound  for the axial-vector mass
reaches its minimum for  $r=2.8$ ($r=3.3$), yielding
$M_A>1.2$~TeV   ($M_A>1.4$~TeV).
It is important to stress that the agreement of the {\it bad\/} solution
with the ``experimental'' $S$ parameter is always reached at the price of
a large NLO correction $\delta_{\rm NLO}^{(1)}$. For any $r>1$,
we find that the  vector mass values compatible with the data
lead to   $(F_{V}^{r\, 2}-F_{A}^{r\, 2})/v^2 \lsim 0$; {\it i.e.},
a very large deviation with respect to the LO WSR,  $F_{V}^2-F_{A}^2=v^2$.
Hence, even though it does not contradict any of the constraints studied in
this subsection, we consider the {\it bad\/} set of solutions  to be clearly disfavoured.

\section{Summary}
\label{sec.conclusions}

We have presented a one-loop calculation of the oblique $S$ parameter within
Higgsless models of EWSB and have analyzed the phenomenological implications
of the available electroweak precision data. We have followed an effective field theory approach, considering a generic Goldstone Lagrangian associated with
the chiral symmetry group $SU(2)_L\times SU(2)_R$, spontaneously broken to $SU(2)_{L+R}$.
Strongly-coupled models of EWSB are characterized by the presence of massive resonance states,
which can be easily incorporated into the effective Lagrangian. We have considered the lightest vector and axial-vector resonances, which couple to the electroweak gauge bosons and the Goldstone bosons, and have written down the most general chiral-invariant Lagrangian, containing up to two resonance fields and no more than two derivatives. We do not include higher-derivative operators because they
would violate the assumed short distance behaviour of the relevant two-point function entering our analysis. The necessary formalism is well-known in low-energy QCD and the results can be easily adapted to the electroweak case with simple notational changes.

Our calculation takes advantage of the dispersive representation of $S$ in terms of
Im$\Pi_{30}(s)$, advocated by Peskin and Takeuchi~\cite{Peskin:92}.
The short-distance operator product expansion guarantees the convergence of the
dispersive integral. Requiring that the correlator computed within the effective
low-energy theory should satisfy the correct high-energy behaviour, it is possible
to perform the dispersive integration and obtain the predicted finite value of $S$.
The dispersive approach avoids all technicalities associated with the renormalization procedure,
allowing us to understand the underlying physics in a much more transparent way.
For instance, one avoids the artifacts from unphysical cut-offs present in previous calculations,
which just manifested the need for local countertems to account for a proper UV completion.
In our approach, the necessary low-energy couplings are determined through short-distance
conditions. Therefore, a crucial ingredient of our calculation is the assumed UV behaviour of the relevant Green functions. We follow closely the procedure developed previously by our group to compute the
low-energy couplings of $\chi$PT at the NLO in $1/N_C$, with $N_C$ the number of QCD colours,
using the R$\chi$T description of the lightest resonances.
The calculation of the electroweak $S$ parameter is analogous to the computation of the $\chi$PT coupling $L_{10}$, performed in Ref.\cite{L10}.

Given our ignorance about the underlying fundamental theory responsible for the EWSB, we can only
provide an approximate description of the true spectral function through a finite number of
contributions. Our parametrization includes the Goldstone pole, the one-particle
exchanges of the lightest vector and axial-vector resonances and the three lowest-mass cuts with two particles, {\it i.e.}, the $\pi\pi$, $V\pi$ and $A\pi$ intermediate states.
While there are certainly many other absorptive contributions, their higher thresholds make them
less relevant at low energies. Moreover, by enforcing the truncated spectral function to satisfy the short-distance behaviour of the full correlator, one expects to obtain a very reliable representation of $\Pi_{30}(s)$ in the relevant region of momentum transfer, specially given its very good convergence properties in the UV. This has been thoroughly tested in QCD with the analogous
correlation function of a left and a right currents\cite{MHA}. When adding more intermediate states, in order to enlarge the range of validity of the spectral representation, the low-energy resonance couplings just adapt their values slightly to accommodate the new contributions without distorting the short-distance fall-off.

The WSRs provide a very powerful constraint to determine the spectral function. In asymptotically-free gauge theories, where the two WSRs are satisfied, they allow us to fix most of the couplings that are relevant for the computation of $S$. 
They require $M_A > M_V$ and at LO force the vector mass to the lower bound $M_V>1.5$~TeV
($M_V>2.3$~TeV) at the $3\sigma$ ($1\sigma$) level.
The smooth UV behaviour of the spectral function, dictated by the WSRs, implies also a small one-loop contribution. At NLO we find that the present experimental value of $S$ requires
that $M_V > 1.8$~TeV (3.8 TeV) and $M_A > 2.5$~TeV (6.6~TeV) at the 3$\sigma$ (1$\sigma$) level.
Higher values are obtained if one requires the VFF and AFF constraints to be also
satisfied; we find in this case $M_V > 2.4$~TeV (6.2 TeV) and $M_A > 4.0$~TeV (10.2~TeV) at the 3$\sigma$ (2$\sigma$) level.
Therefore, in Higgsless electroweak theories
satisfying the two WSRs, such as the usual technicolour models, the associated spectrum of
vector and axial-vector resonances should be much heavier than the 1 TeV scale.

The second WSR is not expected to be fulfilled in Conformal Technicolour models and in most
Walking Technicolour scenarios with non-trivial UV fixed points~\cite{Peskin:92,S-Orgogozo:11}.
However, the first WSR has a much broader range of applicability, including these two last types
of strongly-coupled EWSB theories. We have explored the consequences of dropping the constraints
from the second WSR, with alternative short-distance conditions besides the less constrained case where only the first WSR is imposed. The most important change is that, in the
absence of the second WSR, we can no-longer determine the resonance couplings $F_V^r$ and $F_A^r$, and therefore $S$.
However, assuming that the mass hierarchy $M_V < M_A$ remains still valid, the first WSR provides a lower bound on $S$, which at LO takes the simple form 
$S_{\mathrm{LO}} > 4\pi v^2/M_V^2$.
Adding only the NLO corrections from the two-Goldstone cut this lower bound becomes slightly stronger, requiring $M_V > 1.8$~TeV at the 3$\sigma$ level; the lower bound stays above
the 1$\sigma$ experimental band for any value of $M_V$ (one reaches the closest agreement for $M_V\simeq 5.4$~TeV at the $1.4\sigma$ level).

The full NLO analysis, including also the $V\pi$ and $A\pi$ cuts, is more cumbersome because of the larger number of parameters which, without the second WSR, are less constrained. Imposing the VFF and AFF constraints, plus the first WSR, we have found two possible solutions, which are functions of the resonance masses. One of them obeys
quite well the two LO WSRs,
and leads to a lower bound $M_V> 1.8$~TeV when requiring compatibility with the experimental data at the $3\sigma$ level. 
The other solution violates badly the second
WSR, at LO, for any values of the resonance masses. This {\it bad\/} solution
makes possible to get a lower bound on $S_{\rm NLO}$ in agreement with the experimental data with vector resonance masses below 1 TeV; however, the compatibility is achieved through a very large NLO correction
$\delta_{\rm NLO}^{(1)}$, which implies 
$(F_{V}^{r\, 2}-F_{A}^{r\, 2})/v^2 \lsim 0$; {\it i.e.},
a very large deviation with respect to the LO WSR,  $F_{V}^2-F_{A}^2=v^2$.
Thus, although the presence of light resonance states cannot be generically excluded in strongly-coupled theories where the second WSR is not satisfied, this possibility
requires huge quantum corrections and looks quite unlikely.

Therefore, the $S$ parameter requires a quite high resonance mass scale, beyond the 1 TeV region, in most strongly-coupled scenarios of EWSB at the one-loop level:
$M_V>1.8$~TeV at $3\sigma$. In order to avoid our constraints, one would need
a non-asymptotically-free Higgsless model where the second WSR is violated and, moreover, either the first axial state is lighter than the first vector resonance or (and)
the first WSR receives quantum corrections larger than 100\%.
We are not aware of any interesting model with these properties.

Further constraints on Higgsless electroweak theories can be obtained from the
oblique parameter $T$. Unfortunately, in this case the absence of a known dispersive representation makes more difficult to make a reliable model-independent calculation.
In a future work we plan to investigate which UV assumptions are necessary to perform such a
computation.

{ \bf Note Added} 

During the editorial process, the LHC experiments announced the discovery of a new boson with mass around 125 GeV,   
probably a scalar (although  further analyses are needed to identify its spin). A first estimate of the
impact of a non-standard light scalar on the $S$ and $T$ parameters has also appeared~\cite{Foadi:2012ga}, in the context of strongly-interacting models with a near conformal dynamics. The one-loop contribution to $S$ from a scalar resonance state can be investigated within the same framework we have adopted here. From our previous results in Refs.~\cite{L8,L10}, one could expect a slight decrease of the predicted $S$ parameter and, therefore, the vector resonance mass lower bound. Nevertheless, the numerical impact of the scalar contributions is expected to be moderate, leaving unchanged our main conclusion: the lightest spin-1 resonance mass should remain well above 1 TeV.  Work in this direction is in progress.

\section*{Acknowledgments}

This work has been supported in part by the Spanish Government [grants
FPA2007-60323, FPA2011-23778 and CSD2007-00042 (Consolider Project CPAN)], the
Generalitat Valenciana [Prometeo/2008/069], the Universidad CEU Cardenal Herrera [grants PRCEU-UCH15/10 and PRCEU-UCH35/11] and the MICINN-INFN fund AIC-D-2011-0818.
A.P. would also like to thank the hospitality of the Physics Department of the Technical University of Munich, where this work was started, and the support of the Alexander von Humboldt Foundation.

\appendix

\section{Dispersion relation with poles in the $s$ channel}
\label{app.dispersive}

\begin{figure}[t]
\begin{center}
\includegraphics[angle=0,clip,width=5cm]{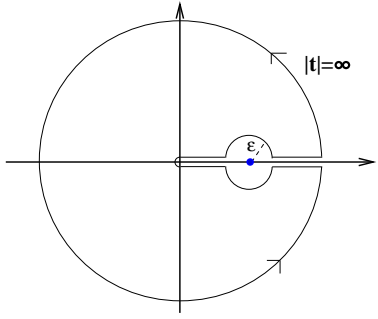}
\caption{\label{fig.circuito}
Integration circuit.}
\end{center}
\end{figure}

The vector and axial-vector resonances have their corresponding poles in the complex plane,
which at LO are located in the real positive axis. A Dyson summation of their self-energies
would generate non-zero resonance widths, moving these poles away from the real axis. However,
in a pure perturbative calculation, such as the one considered here (without Dyson summations),
the resonance poles remain in the real axis and need to be properly taken into account in
the dispersion relations. The one-loop diagrams in Figure~\ref{NLO_graphs} reveal the presence of
single and double poles. Therefore, let us consider a correlation function of the form
\be
\Pi(t)\, =\, \Frac{D(t)}{\left(M_R^2  -  t\right)^2} \, ,
\ee
with $M_R$ a resonance mass and $D(t)$ a function analytic in the whole complex plane
except for the unitarity logarithmic branch (without poles). We will assume that
$|t^{-1}\Pi(t)|\to 0$ when $|t|\to \infty$, so the following once-subtracted relation can be used:
\be
\Pi(s) \, = \, \Pi(0)\, +\, \Frac{s}{2\pi i} \, \oint \  \mathrm{d}t\;\, \Frac{\Pi(t)}{t  (t -  s) }  \, ,
\ee
with the complex integration contour indicated in Figure~\ref{fig.circuito}.

The contribution from the external circle is zero (at infinity radius),
while the integration along the straight lines above and below the real axis result in
the principal part of the usual dispersive integral along the cut, since
$\Pi(s + i\eta) -\Pi(s - i\eta) = 2 i \,\mathrm{Im} \Pi(s)$.
Along the infinitesimal circle $t= M_R^2 + \epsilon \,\mathrm{e}^{i\theta}$, one gets
additional contributions from the massive pole. The final dispersive result can be written
as:
\begin{eqnarray}\label{eq.master}
\Pi(s) & = & \Pi(0)\, -\, \frac{s}{M_R^2}\, \left\{ \Frac{\mbox{Re}\, D'(M_R^2)}{M_R^2-s}  -  \Frac{\mbox{Re}\, D(M_R^2)}{\left(M_R^2-s\right)^2}\right\}
\nonumber\\[5pt]
&+ &\frac{s}{\pi}\;\lim_{\epsilon\to 0} \left\{
\left(\int_0^{M_R^2-\epsilon} + \int_{M_R^2+\epsilon}^\infty \right)
\mathrm{d}t\;\, \Frac{\mbox{Im}\Pi(t) }{t  (t\, -\, s)}
\, - \,
\Frac{2}{\epsilon}
\, \lim_{t\to M_R^2}
\left[(M_R^2-t)^2\, \Frac{\mbox{Im}\Pi(t) }{t (t\,-\,s)}
\right]\right\} ,\quad\;
\end{eqnarray}
with $D'(t)\equiv \Frac{d}{dt}D(t)$.

To determine the correlator one needs the spectral function over the whole cut, as well as
the real part of $D(t)$ and its first derivative at $t=M_R^2$.
This corresponds to providing a renormalization prescription for the corresponding coupling and resonance mass \cite{L10}. The inclusion of additional massive poles can be performed in a straightforward way.

Our computation of the correlator $\Pi_{30}(s)$ includes one vector and one axial-vector poles.
Reabsorbing the pole contributions in the first line of Eq.~(\ref{eq.master}) into a redefinition
of resonance couplings ($F_V^r$, $F_A^r$) and masses ($M_V^r$, $M_A^r$), we can rewrite
the one-loop result in the form given in Eq.~(\ref{eq.T-NLO}) with
\begin{equation}\label{eq.Pibar30}
\overline{\Pi}(s)
 =
\frac{4}{g^2\tan{\theta_W}}\;  \;\lim_{\epsilon\to 0} \left\{
\frac{1}{\pi}\;\int_{\mathcal{I}_\epsilon} \,
\mathrm{d}t\;\, \Frac{\mbox{Im}\Pi_{30}(t) }{t  (t\, -\, s)}
\, - \,
\Frac{2}{\pi\epsilon}
\; \sum_R\; \lim_{t\to M_R^2}\;
\left[(M_R^2-t)^2\, \Frac{\mbox{Im}\Pi_{30}(t) }{t (t\,-\,s)}
\right]\right\}\, ,
\end{equation}
with $\mathcal{I}_\epsilon \equiv (0,\infty)
-\cup_{_{R}} (M_R^2-\epsilon,M_R^2+\epsilon) $.

The corresponding contribution to the electroweak $S$ parameter in Eq.~(\ref{eq.Sbar})
is given by
\be\label{eq.Sbar2}
\overline{S} \, = \, \Frac{16}{g^2 \tan{\theta_W}}\; \;
\lim_{\epsilon\to 0}\left\{ \int_{\mathcal{I}_\epsilon}
\!\!\! \mathrm{d}t\;  \Frac{\rho (t) }{t }
\,-\, \Frac{2}{\epsilon}\, \sum_R\, \lim_{t\to M_R^2}
\left[(M_R^2-t)^2\, \Frac{\rho (t) }{t}
\right] \right\} ,
\ee 
where
$\rho(t)=\mathrm{Im}\widetilde{\Pi}_{30}(t)-$Im$\widetilde{\Pi}_{30}^{\mathrm{SM}}(t)$.

\section{Spectral function contributions}
\label{app.spectralfunctions}

Here we provide the explicit expression of the imaginary part of $\Pi_{30}(s)$. For simplicity we split the spectral function in its different absorptive channels:
\begin{eqnarray}
\mathrm{Im} \Pi_{30}(s) &=&
\frac{g^2 \tan{\theta_W}}{192\pi}\;\left\{
\left.\mathrm{Im} \hat\Pi_{30} (s)\right|_{\pi\pi} \,+\, \left.\mathrm{Im} \hat\Pi_{30}(s) \right|_{A\pi} \, +\, \left.\mathrm{Im} \hat\Pi_{30}(s) \right|_{V\pi} \, +\,\cdots
\right\} .
\end{eqnarray}
We have only considered the contributions from two-particle cuts with two Goldstones or one Goldstone plus one massive resonance.
We have found
\begin{eqnarray}\label{eq.SpecPP}
\left.\mathrm{Im} \hat\Pi_{30}(s) \right|_{\pi\pi} &=& \theta(s)\;
s \; \left[ 1 + \frac{F_VG_V}{v^2} \frac{s}{M_V^2-s} \right]^2 \, ,
\\[10pt]\label{eq.SpecVP}
\left.\mathrm{Im} \hat\Pi_{30}(s) \right|_{V\pi} &=& \theta(s-M_V^2)\; \left( s-M_V^2\right)
\frac{1}{v^2} \,\left\{
  -\frac{2 G_V^2 M_V^4}{s^2}
  -\frac{8 F_V G_V M_V^2}{s} -\frac{F_V^2 M_V^2}{s} \phantom{\Bigg]}
\right. \nn \\ && \left.
    +\, 4 F_V G_V -4 F_V^2 + 6 G_V^2
  +\frac{4 s F_V G_V}{M_V^2} -\frac{s F_V^2}{M_V^2}
    -\frac{4 s G_V^2}{M_V^2}
\right. \nn \\ && \left.
+\, \frac{2 F_A \left(s-M_V^2\right)}{s M_V^2 \left(s-M_A^2\right)}
\bigg[ F_V \left[4 s (\kappa +\sigma ) M_V^2+\sigma  M_V^4
+ s^2 (2 \kappa +\sigma )\right]
\right. \nn \\ && \hskip 3.1cm\left.
 -\, 2 G_V \left(s-M_V^2\right) \left[ (\kappa +2 \sigma ) M_V^2+s
   (2 \kappa +\sigma )\right]\bigg]
\right. \nn \\ && \left.
 -\,\frac{F_A^2 \left(s-M_V^2\right)^2}{s M_V^2 \left(s-M_A^2\right)^2}
  \bigg[2 s \left(\kappa ^2+4 \kappa  \sigma +2 \sigma^2\right) M_V^2
  +\sigma ^2 M_V^4
  +s^2 (2 \kappa +\sigma )^2\bigg]
\right\} ,\quad
\nn\\ &&\\[10pt]\label{eq.SpecAP}
\left.\mathrm{Im} \hat\Pi_{30}(s) \right|_{A\pi} &=& \theta(s-M_A^2)\; \left( s-M_A^2\right)
\frac{1}{s v^2 M_A^2} \,\left\{
   F_A^2  \bigg(4 s M_A^2+M_A^4+s^2\bigg)
\right. \nn \\ && \left.
+\, \frac{2 F_A F_V \left(M_A^2-s\right)}{s-M_V^2} \bigg[ M_A^4
   (2 \kappa +\sigma )+4 s M_A^2 (\kappa +\sigma )+s^2 \sigma \bigg]
\right. \nn \\ && \left.
   +\, \frac{F_V^2 \left(s-M_A^2\right)^2}{\left(s-M_V^2\right)^2}
   \bigg[ M_A^4 (2 \kappa +\sigma )^2+2 s M_A^2 \left(\kappa^2+4 \kappa  \sigma +2 \sigma ^2\right)+s^2 \sigma^2\bigg]    \right\}   \, ,
 \end{eqnarray}
where $\kappa$ and $\sigma$ are the combinations of chiral couplings defined in Eq.~(\ref{eq.kappa}).

In order to study the short-distance constraints dictated by the OPE it is convenient to show the high-energy expansion of $\mathrm{Im}\Pi_{30} (s)$:
\begin{eqnarray}
\mathrm{Im} \Pi_{30}(s) &=& \frac{g^2 \tan{\theta_W}}{192\pi \,M_A^2\, M_V^2 \,v^4}\;\sum_{n=-2} \,\frac{c_{n}}{s^n}\,,
\end{eqnarray}
being
\begin{eqnarray}
c_{-2}&=& v^2 M_V^2 \,\left(F_A-\sigma  F_V\right)^2    -v^2 M_A^2\, \left[F_A\, (2 \kappa +\sigma )-F_V+2 G_V\right]^2 \,, \\ && \nn \\
c_{-1}&=& v^4 M_A^2 M_V^2\, +\, M_A^2 M_V^2 \, \Big\{2 v^2 G_V\, \left[2 F_A\, (5 \kappa +\sigma )-F_V\right] +G_V^2\, \left(F_V^2+10 v^2\right)\nn \\ &&
+\, v^2 \,\left[F_A^2\, \left(10 \kappa ^2+4 \kappa  \sigma -\sigma ^2+3\right)-8 \kappa  F_A F_V+F_V^2\, \left(2 \kappa
   ^2+8 \kappa  \sigma +\sigma ^2-3\right)\right]\Big\} \nn \\ &&
   -\, 2 v^2 F_A M_A^4\, (2 \kappa +\sigma )\, \left[F_A\, (2 \kappa +\sigma )-F_V+2 G_V\right]
   \, +\, 2 \sigma  v^2 F_V M_V^4\, \left(\sigma  F_V-F_A\right) \,, \\ &&\nn \\
c_{0}&=& v^2 M_A^4 M_V^2 \,\Big\{4 F_A \,\left[F_V (3 \kappa +4 \sigma )+G_V (5 \kappa +\sigma
   )\right]\, +\, F_A^2\, \left(20 \kappa ^2+8 \kappa  \sigma -2 \sigma ^2-3\right)\nn \\ &&
   -\, 2 F_V^2 \,\left(\kappa^2+10 \kappa  \sigma +4 \sigma ^2\right)\Big\}
   \, +\, M_A^2 M_V^4 \,\Big\{-2 v^2 G_V \,\left[6 F_A (\kappa
   -\sigma )+7 F_V\right]\nn \\ &&
   +\, v^2 \,\left[F_A^2\, \left(-6 \kappa ^2+12 \kappa  \sigma +8 \sigma ^2\right)-4
   F_A F_V (5 \kappa +4 \sigma )+F_V^2 \left(4 \kappa ^2+16 \kappa  \sigma +2 \sigma
   ^2+3\right)\right]\nn \\ &&
   +\, 2 G_V^2\, \left(F_V^2-3 v^2\right)\Big\}
   -v^2 F_A M_A^6\, (2 \kappa +\sigma )\, \left[3 F_A\, (2 \kappa +\sigma )-2 F_V+4 G_V\right] \nn \\ &&
   +\,\sigma  v^2 F_V M_V^6\, \left(3 \sigma  F_V-2 F_A\right) \,, \\  && \nn \\
c_1 &=& v^2 M_A^6 M_V^2\, \left[4 F_A G_V\, (5 \kappa \!+\!\sigma )+F_A^2 \left(30 \kappa ^2\!+\!12 \kappa  \sigma
   -3 \sigma ^2\!-\!1\right)+2 F_V^2 \left(-3 \kappa ^2\!+\!6 \kappa  \sigma \!+\!4 \sigma ^2\right)\right] \nn \\ &&
   -\, 4 v^2 M_A^4 M_V^4\, \left[3 F_A G_V\, (\kappa -\sigma )+F_A^2\, \left(3 \kappa ^2-6 \kappa  \sigma -4 \sigma
   ^2\right)+F_V^2\, \left(\kappa ^2+10 \kappa  \sigma +4 \sigma ^2\right)\right] \nn \\ &&
   +\, M_A^2 M_V^6\, \Big\{
   v^2 \,\left[F_V^2 \,\left(6 \kappa ^2+24 \kappa
   \sigma +3 \sigma ^2+1\right)-2 F_A^2\, \left(\kappa ^2+10 \kappa  \sigma +4 \sigma
   ^2\right)\right] \nn \\ &&
       -\, 2v^2 G_V\, \left[2 F_A\, (\kappa +5 \sigma )-3 F_V\right]+G_V^2\, \left(3 F_V^2-2 v^2\right)\Big\}
       -2 \sigma  v^2 F_V M_V^8\, \left(F_A-2 \sigma F_V\right) \nn \\ &&
   -\, 2 v^2 F_A M_A^8\, (2 \kappa +\sigma )\,\left[2 F_A\, (2 \kappa +\sigma )-F_V+2 G_V\right] \,. \label{c1}
\end{eqnarray}

The first WSR implies $c_{-2}=c_{-1}=c_0=0$. If moreover the second WSR is fulfilled, the constraint $c_{1}=0$ is used too.


\end{document}